\documentclass[bibyear]{aa}
\usepackage{amsmath,amstext}
\usepackage{natbib,twoopt}
\usepackage[breaklinks=true,colorlinks=true,linkcolor=blue,citecolor=blue,urlcolor=blue]{hyperref}
\usepackage{amssymb}
\usepackage{dblfloatfix}
\newcommand{\Msun}{M_\odot}
\newcommand{\MBH}{M_{\rm BH}}

\begin{document}

\titlerunning{Connecting the little dots in polarized light}

\title{Connecting the little dots in polarized light}

\author{Piero Madau\inst{1,2}
\and
Roberto Maiolino\inst{3,4,5}
\and
Francesco D'Eugenio\inst{3,4}
}
\institute{Dipartimento di Fisica ``G. Occhialini,'' Università degli Studi di Milano-Bicocca, Piazza della Scienza 3, I-20126 Milano, Italy \and Department of Astronomy \& Astrophysics, University of California, 1156 High Street, Santa Cruz, CA 95064, USA \and
Kavli Institute for Cosmology, University of Cambridge, Madingley Road, Cambridge CB3 0HA, UK 
\and Cavendish Laboratory, University of Cambridge, 19 JJ Thomson Avenue, Cambridge CB3 0HE, UK \and
Department of Physics and Astronomy, University College London, Gower Street, London WC1E 6BT, UK}

\abstract{
Recent observations of the local Little Red Dot (LRD) analog SDSS~J1025+1402 have revealed that the optical continuum is polarized at the $\sim1.5\%$ level, while broad H$\alpha$ has lower polarization ($\sim0.7\%$) and a different polarization angle. We investigate whether these properties can be reproduced in a scenario in which LRDs are dust-obscured Little Blue Dots (LBDs) viewed at high inclination and powered by super-Eddington accretion. In this framework, electron scattering in the geometrically thick inner accretion flow produces intrinsic continuum polarization, while unpolarized emission from the outer thin disk dilutes the signal, particularly at optical wavelengths. Broad H$\alpha$, which is emitted outside the compact disk atmosphere, lacks this intrinsic disk-atmosphere polarization. A circumnuclear dust screen then contributes dichroic polarization to both the continuum and broad-line emission, producing the lower line polarization and its different position angle. We show that this model reproduces the main observed polarization properties, although degeneracies between accretion rate and inclination remain. 
Standard equatorial- and polar-scattering AGN polarization models fail to reproduce the observed continuum and broad-line polarization simultaneously.
 These results illustrate the potential of spectropolarimetry as a probe of accretion-flow geometry and orientation in the little-dot population, complementary to spectral-energy-distribution fitting and line diagnostics.
}


\maketitle
\section{Introduction}
\label{sec:intro}

The {James Webb Space Telescope} has revealed a substantial population of compact broad-line active galactic nuclei (BLAGNs) at redshifts $z \gtrsim 4$, powered by accretion onto black holes with inferred masses in the range $\sim 10^{6}$--$10^{8}\,\Msun$ \citep[e.g.,][]{Harikane2023AGN,MaiolinoAGN,Taylor2025_BHMF,Juod2026}. A subset of these objects displays red, V-shaped UV--optical continua and is commonly identified as Little Red Dots (LRDs), while the majority show bluer continua and are classified as Little Blue Dots (LBDs) \citep{Brazzini2026,Geris2026}. In current JWST samples, LRDs account for only $\sim10\%$--30\% of the compact BLAGN population, although the inferred fraction depends on luminosity, redshift, and selection band \citep{Hainline2025,Taylor2025_BHMF,Madau_LF2026}, suggesting that the observed diversity is more likely a statistical property of a single population than evidence for two fundamentally distinct classes \citep{Billand2026}.

In \citet{MadauMaiolino2026}, we proposed that LRDs and LBDs are different inclination-dependent manifestations of the same underlying population of compact, super-Eddington AGNs. In that framework, the central engine consists of a geometrically thick, radiation-supported accretion flow that emits anisotropically, surrounded by a broad-line region (BLR) concentrated toward the equatorial plane and a dusty circumnuclear component with a modest covering factor. Lines of sight at relatively low inclination avoid significant dust attenuation and appear as LBDs, whereas LRDs correspond to the same systems viewed along directions that intercept the dusty component.

This orientation-based picture has recently received two additional lines of support. First, \citet{Madau_LF2026} showed that the observed UV luminosity function of LRDs can be reproduced as the dust-intercepted minority tail of the compact BLAGN parent population, and that the same forward model also predicts the larger inferred black-hole masses of UV-selected LRDs relative to unobscured LBDs as a consequence of selection through dust attenuation. Second, \citet{MadauWings2026} demonstrated that a stratified BLR embedded in this same super-Eddington geometry can account for the extended, nearly exponential Balmer wings observed in both LRDs and LBDs, without invoking electron scattering as the dominant line-broadening mechanism. Additional consistency comes from AGN-powered narrow-line ratios and Ly$\alpha$ emission, which show that ionizing radiation escapes from LRD nuclei, and from broad-Balmer equivalent widths and line ratios that are consistent with the predictions of the dusty, anisotropic model \citep{MadauMaiolino2026,Geris2026,Sok2026}. Together, these results suggest that continuum colors, population demographics, black-hole mass estimates, and broad-line profiles can all be understood within a single rapidly accreting unification framework.

A remaining untested prediction of this scenario concerns the polarization properties of the continuum-emitting accretion flow itself. Because the observed radiation originates from an optically thick, scattering-dominated, highly aspherical surface, the escaping continuum should acquire a non-zero and strongly inclination-dependent linear polarization. We solve the scalar wall-to-wall irradiation of the thick flow, assign the local Chandrasekhar polarization at the final escape surface, and integrate the resulting Stokes fluxes over the visible photosphere. We show that, when combined with a modest dichroic contribution, this geometry can account for the nearly gray continuum polarization of J1025+1402 at the level of $\simeq1.5\%$, as well as the lower polarization of broad H$\alpha$ and its position-angle offset relative to the continuum.

\section{Polarization model}
\label{sec:pol}

We compute the emergent continuum polarization with the geometrical and
radiative framework adopted in \citet{Madau2026}, including its thick-disk
geometry, non-local funnel self-irradiation, and self-shadowing of the hot
inner regions at high inclination. The polarized signal is therefore derived
from the radiation field that shapes both the continuum SED and the BLR
illumination in \citet{MadauMaiolino2026}.

\begin{figure}[!ht]
\centering
\includegraphics[width=\columnwidth]{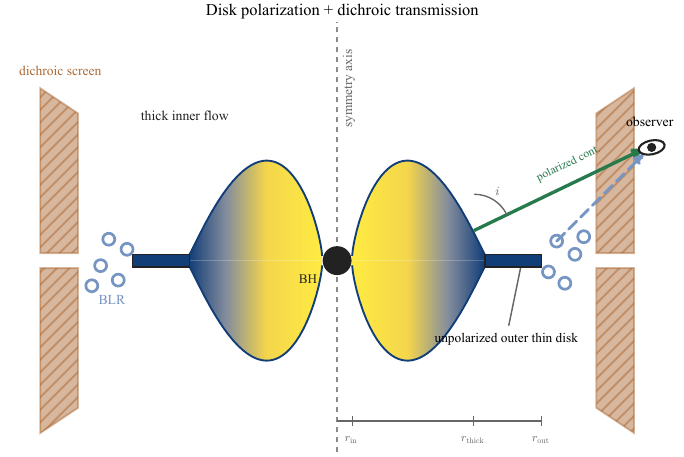}
\caption{Baseline disk-polarization and dichroic-transmission geometry. The
geometrically thick flow extends from $r_{\rm in}$ to $r_{\rm thick}$ and is
matched to an unpolarized thin disk extending to $r_{\rm out}$. At the high
inclination $i$ appropriate to an LRD, the inner funnel is self-obscured and
the visible outer rim supplies most of the polarized disk continuum. The
continuum and intrinsically unpolarized broad H$\alpha$ emission then pass
through a circumnuclear dichroic screen. The drawing is schematic and not to
scale.}
\label{fig:baseline_geometry}
\end{figure}

The baseline geometry is shown in Figure~\ref{fig:baseline_geometry}; the
alternative external-scattering configurations tested below are summarized in
Figure~\ref{fig:model_geometry} of Appendix~\ref{app:model_geometry}.
 
\subsection{Geometry}

The super-Eddington accretion flow is modeled as the composite thick+thin solution described in \citet{Madau2026}: a radiation-supported, non-Keplerian torus extending from the inner radius $r_{\rm in}$ to a transition radius $r_{\rm thick}$, smoothly matched to a standard Keplerian thin disk extending to $r_{\rm out}$. We define the dimensionless accretion rate as $\dot m\equiv\dot M/\dot M_{\rm Edd}$, where $\dot M_{\rm Edd}\equiv10L_{\rm Edd}/c^2$. The funnel height $h(r)$ reaches a maximum at $r=r_{\rm max}$ and decreases again toward the outer transition region.

The emitting funnel surface is described in cylindrical coordinates by the height profile $h(r)$ with slope $h'(r)={\rm d}h/{\rm d}r$, and the corresponding surface element and outward unit normal are
\begin{equation}
{\rm d}\Sigma = r\sqrt{1+h'^2}\,{\rm d}r\,{\rm d}\phi,
\quad
\hat{\boldsymbol{n}}=
\frac{(-h'\cos\phi,\,-h'\sin\phi,\,1)}{\sqrt{1+h'^2}}.
\end{equation}
For an observer at inclination $i$ from the symmetry axis, the line-of-sight unit vector is $\hat{\boldsymbol{k}}=(\sin i,\,0,\,\cos i)$ and $\mu(r,\phi;i)\equiv\hat{\boldsymbol n}\mathbin{\cdot}\hat{\boldsymbol k}$. Only surface elements with $\mu>0$ and not blocked by self-shadowing contribute to the observed flux, using the same visibility conditions as in \citet{Madau2026}.
 
\subsection{Radiative transfer on the funnel wall}

At the high accretion rates and temperatures characteristic of
super-Eddington flows, the funnel photosphere is highly ionized. Its
UV/optical free--free and bound--free opacities are consequently negligible
compared with electron scattering \citep{Madau2026}. The outgoing continuum
therefore acquires net linear polarization through the angular asymmetry of
the final electron scatterings near the photosphere. In addition to the
locally generated flux, each surface element receives radiation from other
visible parts of the funnel. Because the wall atmosphere has
single-scattering albedo close to unity at UV/optical wavelengths, this
intercepted flux is not thermally absorbed and re-emitted as blackbody
radiation. Instead, it is redistributed by multiple electron scatterings
within the local wall atmosphere before contributing to the outgoing
radiation field. The appropriate description is therefore a self-consistent
multiple-scattering problem on the funnel surface, not a single-scatter
source term.

We discretize the axisymmetric funnel wall into radial rings indexed by \(k\). Following \citet{Madau2026}, the total outgoing spectral flux associated with ring \(k\) satisfies
\begin{equation}
\mathcal F_{\nu,k} = \mathcal F_{\nu,k}^{\rm local}
+ \frac{1}{\pi}\sum_j B_{kj}\,\mathcal F_{\nu,j},
\label{eq:Frad}
\end{equation}
where $\mathcal F_{\nu,k}^{\rm local}$ is the locally generated flux and the geometric kernel
\begin{align}
B_{kj} &= \int_0^{2\pi}
\frac{\bigl[-\hat{\boldsymbol{n}}_k\cdot\hat{\boldsymbol{L}}\bigr]_+\,
      \bigl[\hat{\boldsymbol{n}}_j\cdot\hat{\boldsymbol{L}}\bigr]_+}
     {|\boldsymbol{L}|^2}
\,{\rm d}\Sigma_j(\phi), \notag\\[4pt]
{\rm d}\Sigma_j &= r_j\sqrt{1+h_j'^{\,2}}\,\Delta r\,{\rm d}\phi,
\end{align}
is the dimensionless geometric coupling factor, such that \(B_{kj}/\pi\) is the fraction of the flux emitted by ring \(j\) that is intercepted by ring \(k\). Here \(\Delta r\) is the radial ring width, \(\boldsymbol L\) points from a surface element at azimuth \(\phi\) on ring \(j\) to a reference element on ring \(k\), and $[x]_+\equiv\max(x,0)$ restricts the integral to mutually visible surface elements. We solve the linear system in Equation~(\ref{eq:Frad}) by lower--upper (LU) matrix decomposition, which factorizes its coefficient matrix into lower- and upper-triangular matrices, thereby summing all orders of internal wall-to-wall scattering.

For wall-to-wall coupling we adopt the Lambertian closure $I_\nu=\mathcal F_\nu/\pi$. Equation~(\ref{eq:Frad}) consequently propagates only the total outgoing spectral flux integrated over direction; it carries neither the incident angular distribution nor the Stokes components $Q$, $U$, and $V$ between wall encounters. After solving for $\mathcal F_{\nu,k}$, we introduce the observer-direction dependence and local linear polarization using the Chandrasekhar solution for a semi-infinite conservative electron-scattering atmosphere \citep{Chandrasekhar1960}. Thus each wall encounter preserves the incident energy through its contribution to $\mathcal F_{\nu,k}$ but discards the incident angular and Stokes information; a new local Chandrasekhar polarization is assigned at the final escape surface. This closure cannot retain depolarization or position-angle rotation accumulated across successive wall encounters. Appendix~\ref{app:stokes_benchmark} therefore provides a controlled full-Stokes Monte Carlo test of the closure in a simplified funnel, rather than a full transfer calculation for the numerical disk geometry.

With these prescriptions, the wall solve determines the total scalar outgoing flux \(\mathcal F_\nu(r)\), including both locally generated radiation and energy returned by other parts of the funnel. The angular dependence of the radiation reaching the observer is then written as
\begin{align}
I_\nu(r,\mu) &= \frac{\mathcal F_\nu(r)}{2\pi}\,{\cal H}(\mu),
\label{eq:Ilimb}\\[4pt]
Q_\nu(r,\mu) &= \frac{\mathcal F_\nu(r)}{2\pi}\,p(\mu)\,{\cal H}(\mu),
\label{eq:Qlimb}
\end{align}
where \({\cal H}(\mu)\simeq 0.85+1.725\,\mu\) is the limb-darkening function, normalized so that \(\int_0^1 {\cal H}(\mu)\mu\,{\rm d}\mu=1\),\footnote{Our calculations implement the exact Chandrasekhar \(H\)-function solution; the approximate expression for \({\cal H}(\mu)\) is given here only for illustration.}
and \(p(\mu)\) is the Chandrasekhar polarization fraction. Within this plane-parallel, semi-infinite, conservative-atmosphere solution, the latter vanishes for normal emergence (\(\mu=1\)) and rises to \(11.7\%\) at grazing emergence (\(\mu\rightarrow0\)). The \(11.7\%\) value is therefore the maximum of the local plane-parallel Milne solution, rather than a universal upper bound for radiation emerging from a curved, mutually irradiating funnel. We use it only as part of our local closure, whose accuracy is tested in Appendix~\ref{app:stokes_benchmark}. Both functions are evaluated at the observer angle \(\mu=\mu(r,\phi;i)\).

We neglect circular polarization, as appropriate for cold-electron Thomson scattering without incident circular polarization or magnetic propagation effects \citep{Chandrasekhar1960,GoosmannGaskell2007}. We place the observer in the \(xz\)-plane and use the projected disk rotation axis as the reference direction for Stokes \(Q_\nu\). 
Reflection symmetry about this plane pairs elements at \((r,\phi)\) and \((r,-\phi)\) with equal Stokes \(Q_\nu\) contributions and opposite Stokes \(U_\nu\) contributions, where the axes defining \(U_\nu\) are rotated by \(45^\circ\) relative to those defining \(Q_\nu\). The \(U_\nu\) contributions therefore cancel pairwise in the azimuthal integration, leaving the net Stokes vector purely \(Q_\nu\) in the symmetry-aligned frame. Rotation to the celestial reference frame then generally produces nonzero observed \(Q_\nu\) and \(U_\nu\), without breaking the underlying axial symmetry. This cancellation is illustrated with a conical-funnel model in Appendix~\ref{app:conical_toy}.

\subsection{Observed Stokes parameters}

The observed Stokes fluxes are obtained by integrating the specific intensities from Equations~(\ref{eq:Ilimb})--(\ref{eq:Qlimb}) over the visible photospheric surface of the geometrically thick flow. For each surface element at \((r,\phi)\), the local linear-polarization direction is perpendicular to the plane defined by the surface normal and the observer's line of sight. We denote its position angle relative to the projected disk rotation axis by \(\chi(r,\phi;i)\). If \(\psi_{\rm axis}\) is the position angle of that axis relative to the fixed celestial Stokes-\(Q\) direction, the corresponding polarization angle in the observer's reference frame is
\begin{equation}
\chi_{\rm sky}(r,\phi;i)
=
\chi(r,\phi;i)+\psi_{\rm axis}.
\end{equation}
The axisymmetric calculation does not determine \(\psi_{\rm axis}\). In the fits below, we instead use the celestial polarization position angle of the integrated disk component, \(\psi_{\rm disk}\), directly as a nuisance parameter.

The projected area element is \(\mu\,{\rm d}\Sigma\), and the integration extends over all surface elements with \(\mu>0\) that are not blocked by self-shadowing. The celestial-frame Stokes fluxes are then
\begin{align}
f_{\nu,{\rm obs}}
&=
\frac{1}{D^2}
\int_{\Sigma_{\rm vis}}
I_\nu(r,\mu)\,\mu\,{\rm d}\Sigma,
\label{eq:Fnu}\\[4pt]
Q_{\nu,{\rm sky}}
&=
\frac{1}{D^2}
\int_{\Sigma_{\rm vis}}
Q_\nu(r,\mu)\cos 2\chi_{\rm sky}\,
\mu\,{\rm d}\Sigma,
\label{eq:Qnu}\\[4pt]
U_{\nu,{\rm sky}}
&=
\frac{1}{D^2}
\int_{\Sigma_{\rm vis}}
Q_\nu(r,\mu)\sin 2\chi_{\rm sky}\,
\mu\,{\rm d}\Sigma.
\label{eq:Unu}
\end{align}
Here \(D\) is the distance to the source. The quantity \(Q_\nu(r,\mu)\) inside the integrals is the local polarized specific intensity defined by Equation~(\ref{eq:Qlimb}), whereas \(Q_{\nu,{\rm sky}}\) and \(U_{\nu,{\rm sky}}\) are the surface-integrated Stokes fluxes. Similarly, \(\mathcal F_\nu(r)\) is the local outgoing surface flux determined by Equation~(\ref{eq:Frad}), whereas \(f_{\nu,{\rm obs}}\) is the corresponding surface-integrated observed flux. 

\subsection{Polarization dilution by the outer thin disk}

The outer standard thin disk is included in the calculation but treated as unpolarized, contributing only flux that dilutes the polarized thick-flow signal. The total observed flux density is therefore
\begin{equation}
f_{\nu,{\rm obs}}^{\rm tot}=f_{\nu,{\rm obs}}+f_{\nu,{\rm obs}}^{\rm thin},
\end{equation}
whereas \(Q_{\nu,{\rm sky}}\) and \(U_{\nu,{\rm sky}}\) receive no thin-disk contribution. The polarization fraction predicted by the accretion-disk model is consequently
\begin{equation}
p_{\rm disk}(\nu)
=
\frac{
\sqrt{Q_{\nu,{\rm sky}}^2+U_{\nu,{\rm sky}}^2}
}{
f_{\nu,{\rm obs}}+f_{\nu,{\rm obs}}^{\rm thin}
}.
\end{equation}
Because the polarized amplitude is invariant under rotations of the Stokes reference frame, the same polarization fraction is obtained in the symmetry-aligned frame, where the surface-integrated Stokes \(U_\nu\) vanishes.

The assumption of negligible thin-disk polarization is motivated by the
properties of its atmosphere. In the cooler Keplerian regime, absorption
opacity becomes important and the classical \citet{Chandrasekhar1960}
solution for a pure electron-scattering atmosphere no longer applies.
Absorption can either suppress or enhance the polarization, depending on the
thermal source-function gradient, whereas magnetic Faraday rotation can
reduce it further \citep{Laor1990,Koratkar1999,Agol1998,Taverna2021}.
Detailed atmosphere calculations generally predict weak, model-dependent
optical/UV polarization, typically at or below the percent level
\citep{Laor1990,Agol1998,Taverna2021}. We therefore
retain the thin-disk continuum flux but neglect its intrinsic Stokes fluxes,
with possible Faraday depolarization providing additional motivation. This
approximation is based on atmospheric physics, not axisymmetry, since an
inclined axisymmetric scattering disk can produce finite linear polarization.

\begin{figure}[!ht]
\centering
\includegraphics[width=0.9\hsize,trim=0 0 0 14mm,clip]{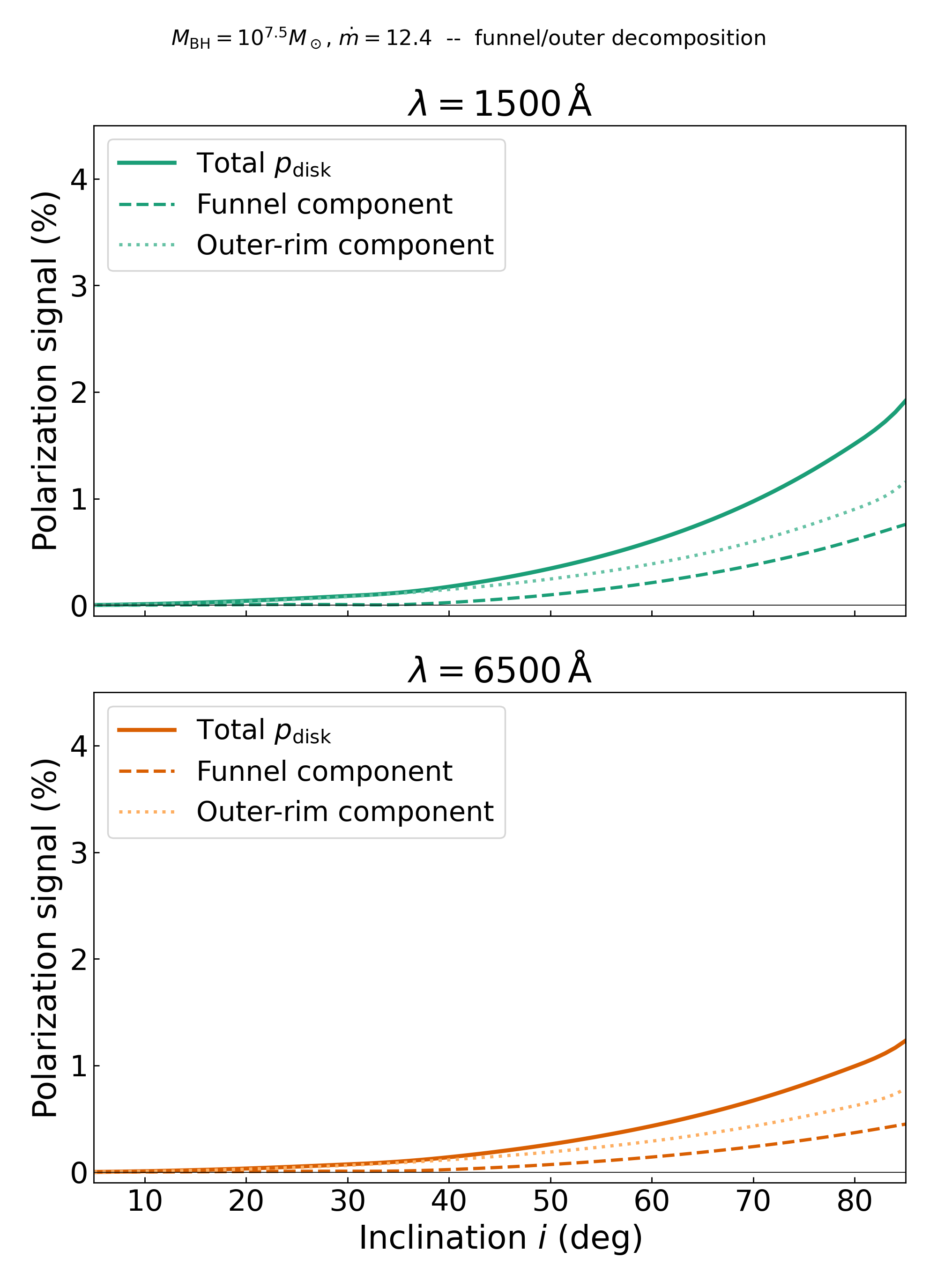}
\caption{Predicted continuum polarization versus inclination for
$\MBH=10^{7.5}\,\Msun$ and $\dot m=12.4$ at rest-frame $1500\,$\AA\ (top)
and $6500\,$\AA\ (bottom). Solid curves show $p_{\rm disk}$; dashed and
dotted curves show the signed inner-funnel and outer-rim contributions to
$Q_{\nu,{\rm sky}}/f_{\nu,{\rm obs}}^{\rm tot}$. The unpolarized thin disk
dilutes the signal, especially at longer wavelengths.}
\label{fig:pol_decomp}
\end{figure}

The unpolarized thin-disk flux lowers the net polarization fraction, particularly at optical wavelengths where its relative contribution is larger. The resulting wavelength-dependent dilution produces a mild UV-to-optical decline in \(p_{\rm disk}(\lambda)\).
Figure~\ref{fig:pol_decomp} illustrates this effect for a fiducial super-Eddington accretor with
\(\MBH=10^{7.5}\,\Msun\) and \(\dot m=12.4\). The polarized thick-flow emission is divided into the inner funnel
(\(r_{\rm in}\leq r\leq r_{\rm max}\)) and the outer rim
(\(r_{\rm max}<r\leq r_{\rm thick}\)); the standard thin disk (\(r_{\rm thick}<r\leq r_{\rm out}\)) contributes only diluting flux. At \(i\lesssim60^\circ\), azimuthal cancellation keeps the net polarization below \(\sim1\%\). Toward edge-on orientations, self-shadowing suppresses the inner funnel and restricts the visible thick-flow surface to a smaller portion of the outer rim. Oppositely signed azimuthal contributions then cancel less completely, so the outer rim dominates the observed polarized flux. At \(i=80^\circ\), the polarization
reaches \(1.5\%\) at \(1500\,\)\AA\ and \(1.0\%\) at \(6500\,\)\AA; the difference reflects the increasing thin-disk contribution toward longer wavelengths.

\begin{figure}[!hbt]
\centering
\includegraphics[width=0.9\hsize,trim=0 0 0 14mm,clip]{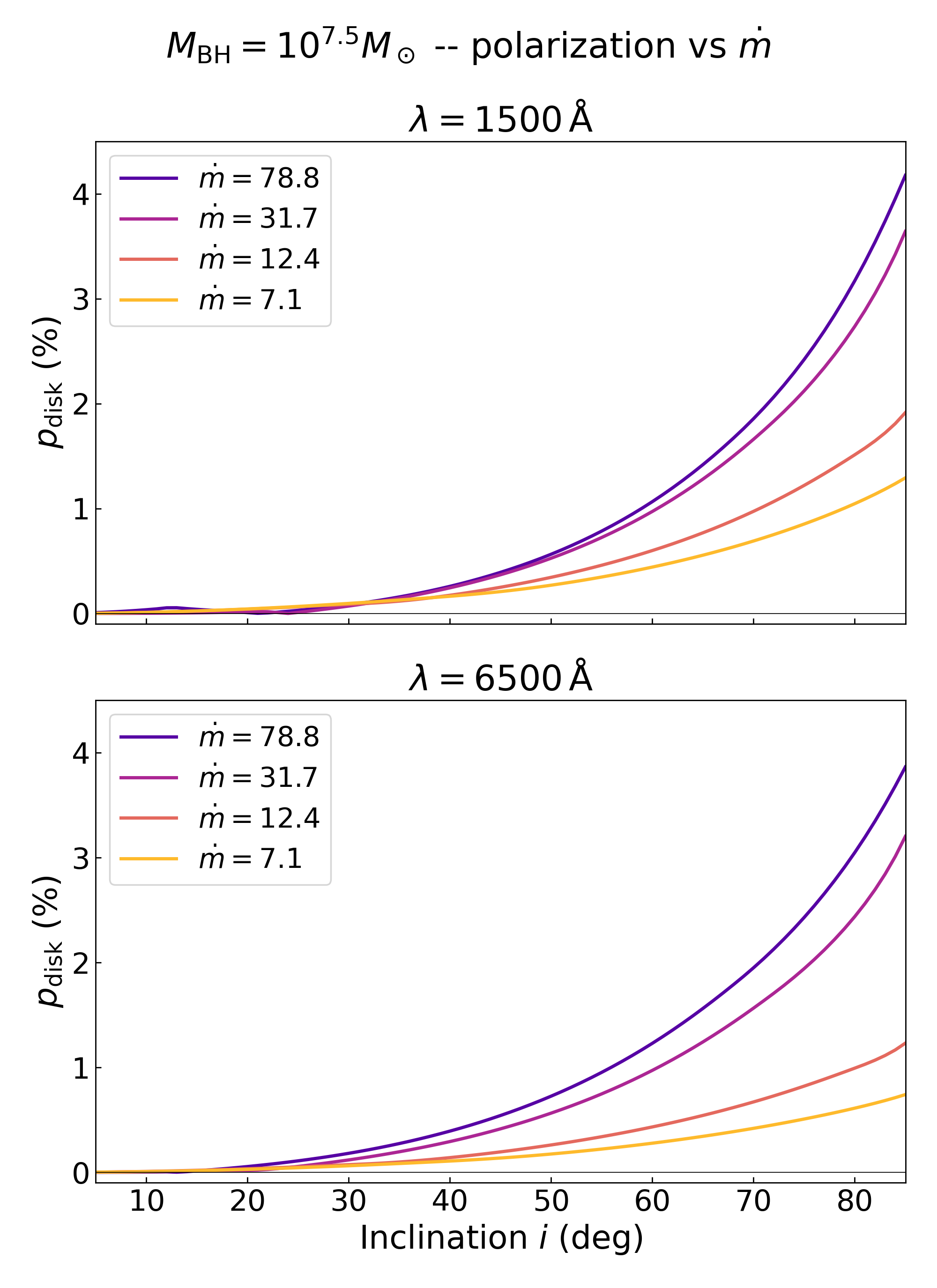}
\caption{Net disk polarization versus inclination for
$\MBH=10^{7.5}\,\Msun$ and several accretion rates, at rest-frame
$1500\,\text{\AA}$ (top) and $6500\,\text{\AA}$ (bottom). Polarization
generally increases with inclination and $\dot m$; thin-disk dilution makes
the UV-to-optical decline stronger at lower $\dot m$.}
\label{fig:pol_grid}
\end{figure}

\begin{figure}[!hbt]
\centering
\includegraphics[width=0.9\hsize,trim=0 0 0 0,clip]{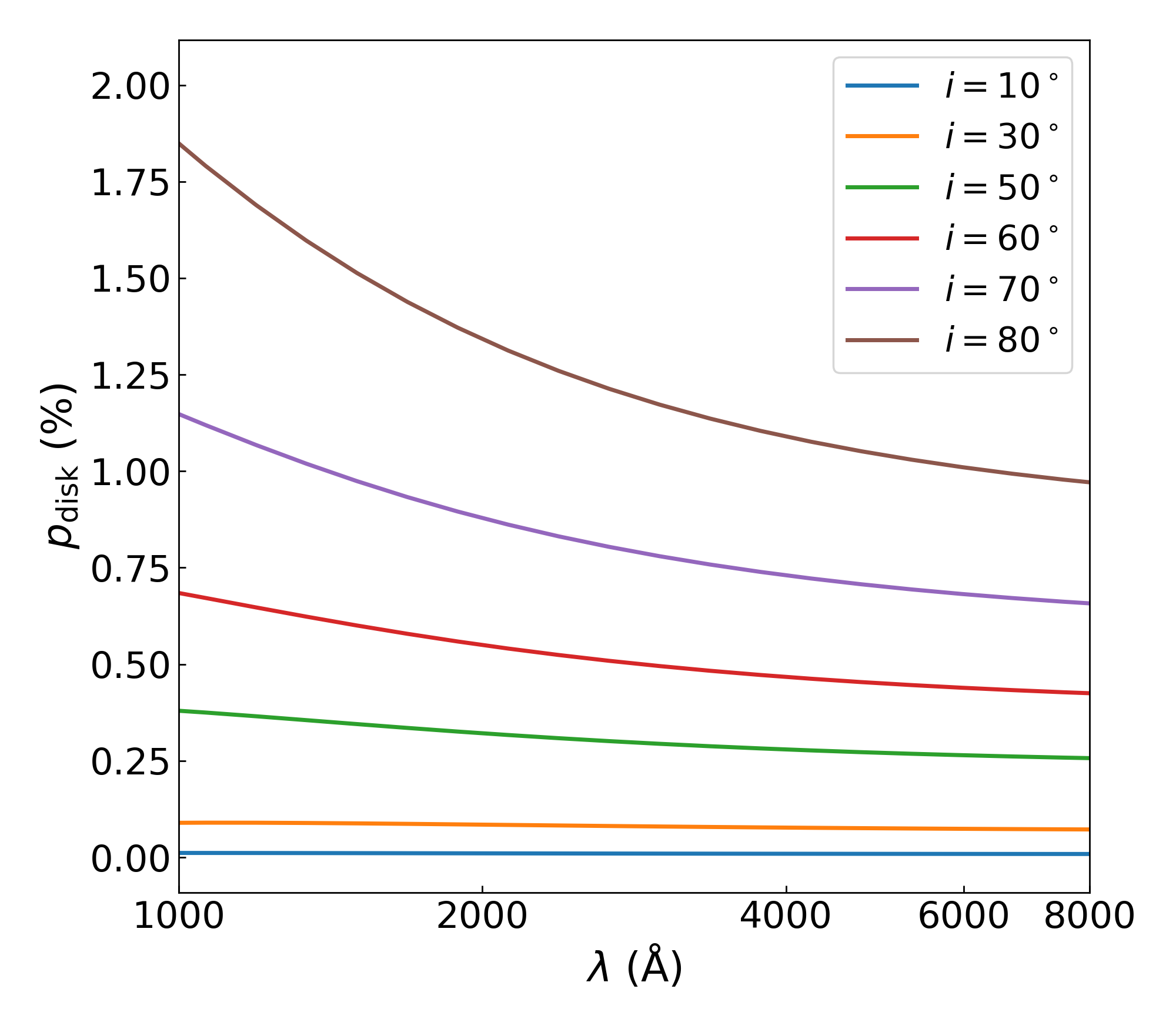}
\caption{Net disk polarization spectrum for $\MBH=10^{7.5}\,\Msun$,
$\dot m=12.4$, and several inclinations. The polarization increases with
inclination and declines toward longer wavelengths as the unpolarized thin
disk contributes a larger fraction of the flux.}
\label{fig:pol_spectrum}
\end{figure}

\subsection{Dependence on accretion rate and wavelength}

The dependence of \(p_{\rm disk}\) on \(\dot m\) is shown in Figure~\ref{fig:pol_grid}. The polarization generally rises with inclination, apart from small non-monotonic features at low inclination where azimuthal cancellation is nearly complete. At \(i=85^\circ\), \(p_{\rm disk}(1500\,\text{\AA})\) reaches \(4.2\%\), \(3.6\%\),
\(1.9\%\), and \(1.3\%\) for \(\dot m=78.8\), \(31.7\), \(12.4\), and \(7.1\), respectively.
The dependence on \(\dot m\) is governed primarily by the extent of the unpolarized outer thin disk rather than by the geometry of the geometrically thick flow alone. At higher accretion rates, \(r_{\rm thick}\) is larger and the thin disk accounts for a smaller fraction of the total optical flux, leaving the thick-flow polarization less diluted. At lower accretion rates, \(r_{\rm thick}\) decreases and the thin
disk extends over a wider radial range, so its unpolarized flux increasingly dilutes the optical polarization. Consequently, \(p_{\rm disk}\) increases with \(\dot m\) at moderate and high inclinations.

The wavelength dependence also varies with \(\dot m\). Comparing the two panels of Figure~\ref{fig:pol_grid}, at \(i=85^\circ\) we find
\begin{equation}
\frac{p_{\rm disk}(1500\,\text{\AA})}
     {p_{\rm disk}(6500\,\text{\AA})}
=
1.08,\ 1.14,\ 1.56,\ \text{and}\ 1.74
\end{equation}
for \(\dot m=78.8\), \(31.7\), \(12.4\), and \(7.1\), respectively. Thus the two higher-rate models are nearly gray, whereas the lower-rate models show a progressively stronger UV-to-optical decline owing to thin-disk dilution. Figure~\ref{fig:pol_spectrum} shows the full wavelength dependence for the fiducial \(\dot m=12.4\) model.

\section{Case study: the LRD analog SDSS~J1025+1402}

Recent optical spectropolarimetry of the nearby LRD analog SDSS~J1025+1402 provides direct constraints on the continuum-emitting region of an LRD-like source \citep{Deugenio2026_pol}. Located at $z\simeq0.1$, J1025+1402 shares the defining properties of high-redshift LRDs, including a compact morphology, a V-shaped UV--optical spectral energy distribution, broad Balmer emission lines with extended non-Gaussian wings, a red optical continuum, and extreme X-ray weakness
\citep{Lin2026,Ji2026}. Its proximity enables signal-to-noise levels
currently inaccessible for the high-redshift population.

The Gemini/GMOS slit image has a seeing FWHM of $0.64''$; a
Moffat-plus-background fit \citep{Moffat1969} leaves no detectable spatial
residual from a host galaxy and constrains the optical effective radius to
$R_{\rm e}<0.32''$, or $R_{\rm e}<620\,$pc \citep{Ji2026}. Thus the rest-frame optical continuum of J1025+1402 is unresolved on sub-kpc scales and appears to be dominated by a single compact source, and we neglect host-galaxy dilution. For illustration, an unpolarized host contribution of $10\%$ to the observed optical luminosity would dilute the polarization by a factor of $0.9$; correcting for it would increase the intrinsic polarization inferred for the compact source by only a factor $1/0.9$, or approximately $11\%$, without changing its polarization angle.

Using VLT/FORS2 optical spectropolarimetry, \citet{Deugenio2026_pol} measured a nearly gray continuum polarization of $p_{\rm cont}=(1.53\pm0.04_{\rm rand}\pm0.20_{\rm syst})\%$ over the rest-frame wavelength range $5300$--$6650\,$\AA. After continuum subtraction, broad H$\alpha$ is less polarized, with $p_{{\rm H}\alpha}=0.58$--$0.84\%$, depending on the estimator. Its polarization angle differs significantly from that of the continuum, with
\begin{multline}
\Delta\vartheta
\equiv
\vartheta_{{\rm H}\alpha}-\vartheta_{\rm cont}
\\
= -48^\circ\pm4^\circ\,({\rm stat})
\pm12^\circ\,({\rm epoch\ systematic}).
\end{multline}
This position-angle offset motivates the two-component polarization model developed below; the fit itself uses the Stokes fluxes directly (Section~\ref{sec:J1025_fit}).

\subsection{Model}

We model the normalized continuum Stokes vector as the sum of two contributions: intrinsic scattering polarization from the super-Eddington flow and dichroic polarization from aligned dust grains in a dusty screen along the LRD sightline. We associate this screen with the circumnuclear dusty component of the LRD--LBD unification model, although the polarization calculation applies equally to any intervening dust-bearing structure along the line of sight. We define the normalized linear Stokes parameters as
\begin{equation}
q(\lambda)\equiv\frac{Q(\lambda)}{I(\lambda)},
\qquad
u(\lambda)\equiv\frac{U(\lambda)}{I(\lambda)},
\end{equation}
so that $p=(q^2+u^2)^{1/2}$. A component with polarization fraction
$p$ and celestial position angle $\psi$ contributes $q=p\cos2\psi$ and
$u=p\sin2\psi$. All position angles are understood modulo $180^\circ$. At
continuum wavelengths,
\begin{align}
q_{\rm cont}(\lambda)
&=
p_{\rm disk}(\lambda)\cos 2\psi_{\rm disk}
+
p_{\rm dic}(\lambda)\cos 2\psi_{\rm dic},
\label{eq:J1025_qobs}
\\
u_{\rm cont}(\lambda)
&=
p_{\rm disk}(\lambda)\sin 2\psi_{\rm disk}
+
p_{\rm dic}(\lambda)\sin 2\psi_{\rm dic},
\label{eq:J1025_uobs}
\end{align}
where $p_{\rm disk}(\lambda)$ is the polarization fraction predicted by the disk model, including dilution by the unpolarized outer thin disk, at inclination $i$ and accretion rate $\dot m$. The quantities $p_{\rm dic}(\lambda)$ and $\psi_{\rm dic}$ describe the dichroic component, while $\psi_{\rm disk}$ is the celestial polarization position angle of the disk component. Equations~(\ref{eq:J1025_qobs}) and (\ref{eq:J1025_uobs}) use the first-order, small-polarization limit and neglect terms of order $p_{\rm disk}p_{\rm dic}$.

For compactness, below we use
$q_{\rm disk}\equiv p_{\rm disk}\cos2\psi_{\rm disk}$,
$u_{\rm disk}\equiv p_{\rm disk}\sin2\psi_{\rm disk}$,
$q_{\rm dic}\equiv p_{\rm dic}\cos2\psi_{\rm dic}$, and
$u_{\rm dic}\equiv p_{\rm dic}\sin2\psi_{\rm dic}$.

We assume that the extended broad-line emission carries negligible intrinsic or scattered polarization and passes through the same dichroic screen as the continuum. Broad H$\alpha$ therefore carries only the dichroic Stokes vector.

The best-fitting inclination, $i\simeq70^\circ$, is near-equatorial, but the
model does not place an opaque classical torus along the line of sight.
Instead, it requires a partially obscured Type~1 view in which the continuum
source and most of the Balmer-emitting BLR are seen directly through a
moderate column of circumnuclear dust. The adopted $A_V\simeq2$--4 mag is
large enough to redden the source but small enough for broad Balmer emission
to remain directly visible through the dusty screen. In classical
low-redshift unification, comparable near-equatorial sightlines generally
intercept larger dust columns and therefore produce Type~2 spectra
\citep{Marin2014}. J1025+1402, however, has a measured gas-phase metallicity
of $Z\simeq0.11\,Z_\odot$ \citep{Ji2026}, making a reduced dust-to-gas ratio
plausible and allowing only moderate extinction even at this inclination.

We interpret this screen as a compact or patchy
circumnuclear structure that covers the continuum source and BLR along our
line of sight but not most of the spatially extended narrow-line region. The
measured narrow-line ratio is
$\mathrm{H}\alpha/\mathrm{H}\beta=3.31\pm0.01$, corresponding to only
$A_V=0.38\pm0.02$ mag; the SDSS spectrum independently gives
$A_V\simeq0.4$--$0.6$ mag \citep{Ji2026}. These values are far below the
$A_V\simeq2$--4 mag relevant to the compact nuclear sightline, so the directly
observed narrow-line emission cannot traverse the same dust column. It must
originate predominantly outside the screen or reach us through lower-column
sightlines. Section~\ref{sec:J1025_core} therefore treats the direct narrow
lines separately.

If $f_{\rm cont}(\lambda)\equiv I_{\rm cont}(\lambda)/I_{\rm obs}(\lambda)$
is the continuum fraction measured from the observed intensity spectrum, the model across H$\alpha$ becomes
\begin{align}
q_{\rm mod}(\lambda)
&=
q_{\rm dic}(\lambda)
+
f_{\rm cont}(\lambda)q_{\rm disk}(\lambda),\\
u_{\rm mod}(\lambda)
&=
u_{\rm dic}(\lambda)
+
f_{\rm cont}(\lambda)u_{\rm disk}(\lambda).
\label{eq:J1025_line_mixing}
\end{align}
Outside the line window, $f_{\rm cont}=1$, and these expressions
reduce to the continuum model above. The retained bins across the
broad-line profile therefore constrain the relative amplitudes and
orientations of the disk and dichroic components; the
continuum-subtracted H$\alpha$ polarization fraction and position
angle are not included as separate summary constraints. This baseline fit uses the observed total-intensity spectrum $I_{\rm obs}(\lambda)$ and its fitted continuum $I_{\rm cont}(\lambda)$ to determine $f_{\rm cont}(\lambda)$ and hence
the line dilution. A fully forward model could instead predict the H$\alpha$ luminosity and profile through a photoionization calculation. Any additional scalar foreground attenuation multiplies all Stokes fluxes by the same wavelength-dependent factor and therefore cancels from the normalized Stokes parameters. This is a transmission geometry: dust scattering would instead require off-axis photons to be redirected into the line of sight and is not included in the broad-line model.

The disk polarization $p_{\rm disk}(\lambda)$ is nearly gray across
the limited optical wavelength range of the observations. Although
electron scattering itself is wavelength-independent, the
disk-integrated polarization retains a weak wavelength dependence
through the radial weighting of the emission and dilution by the
outer thin disk. For the best-fitting disk model discussed below,
$p_{\rm disk}$ changes by only $\simeq0.01$ percentage points between
$5300$ and $6650\,$\AA. Any appreciable residual slope in the modeled
continuum polarization is therefore governed primarily by the
dichroic component, for which we adopt the Serkowski law
\citep{Serkowski1975},
\begin{equation}
p_{\rm dic}(\lambda)
=
p_{\rm max}
\exp\!\left[
-K\ln^2\!\left(\frac{\lambda_{\rm max}}{\lambda}\right)
\right],
\label{eq:J1025_Serkowski}
\end{equation}
where the amplitude $p_{\rm max}$ is fitted directly. We fix $\lambda_{\rm max}=5500\,$\AA, the canonical value for diffuse Galactic sightlines \citep{Serkowski1975}, and set $K=0.92$ using the empirical relation
$K\simeq 1.66(\lambda_{\rm max}/\mu{\rm m})+0.01$ \citep{Whittet1992}.

We verified that the results are insensitive to the adopted
dichroic wavelength dependence. Varying $\lambda_{\rm max}$ from
$4000$ to $7000\,$\AA, with $K$ tied through the Whittet relation,
changes the fitted $p_{\rm max}$ by less than $20\%$ and the minimum
$\chi^2$ by less than $0.5$, while leaving the inferred disk
polarization angle essentially unchanged. This insensitivity arises
because the Serkowski curve is nearly achromatic across the narrow
optical baseline, so its shape cannot be constrained by the present
data.

\begin{figure*}
\centering
\includegraphics[width=0.9\hsize,trim=0 0 0 8.5mm,clip]{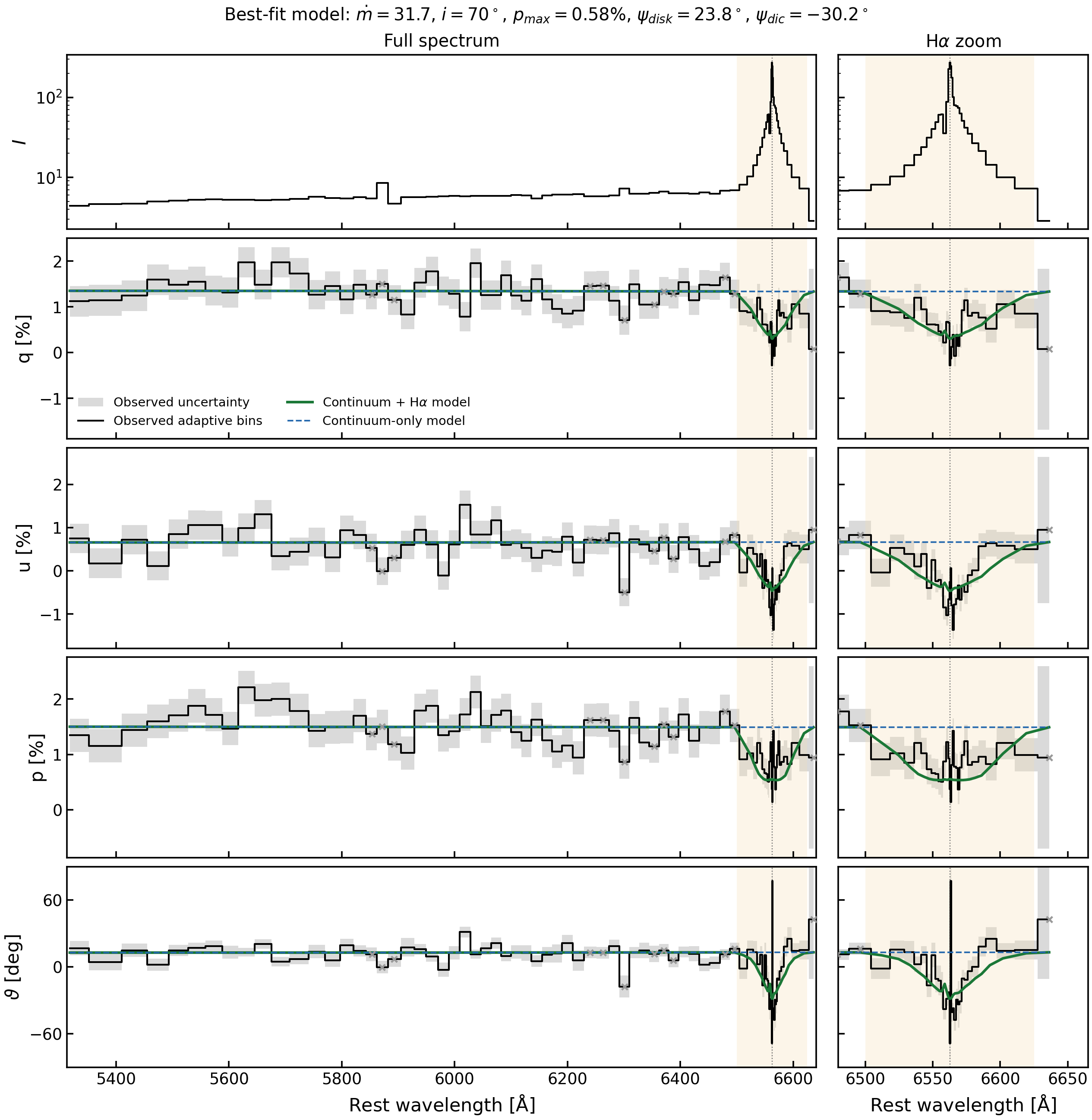}
\caption{Full-spectrum fit to SDSS~J1025+1402 over the fitted range (left)
and around broad H$\alpha$ (right). Rows show the intensity used to determine
$f_{\rm cont}$, normalized Stokes $q$ and $u$, the derived $p$, and the
polarization position angle $\vartheta$. Black steps and gray bands are the FORS2 data and $1\sigma$
uncertainties; green curves show the full model and dashed blue curves the
continuum-only prediction. Shading marks the fitted line window, and gray
crosses mark the masked bins. The model predicts a smooth position-angle
change as the disk-polarized continuum is diluted relative to the dichroically
polarized line. The sharp central excursion occurs only in the data, where
$p$ is small and the position angle is poorly constrained. Only retained $q$
and $u$ values are shown here, but the likelihood is evaluated directly in
$Q$ and $U$ flux. The fit has nominal
$\chi^2/{\rm dof}=210.8/133=1.59$.}
\label{fig:pol_fit}
\end{figure*}

\subsection{Fit and results}
\label{sec:J1025_fit}

In our baseline fit, we do not decompose H$\alpha$ into broad and narrow
components. The complete observed line profile is assumed to pass through
the same dichroic screen. Residual structure near line center may therefore
reflect a narrow-line contribution with polarization different from that of
the broad line, or absorption-related
effects, neither of which is included in our model
\citep[but see][]{Deugenio2026_pol}. We fit the directly measured,
adaptively binned Stokes-flux vectors $\boldsymbol Q$ and $\boldsymbol U$
provided by D'Eugenio et al., together with their full wavelength-bin
covariance matrices $C_Q$ and $C_U$. The modeled fluxes are
equivalently
\begin{multline}
Q_{\rm mod}=I_{\rm obs}q_{\rm dic}+I_{\rm cont}q_{\rm disk},
\\
U_{\rm mod}=I_{\rm obs}u_{\rm dic}+I_{\rm cont}u_{\rm disk}.
\end{multline}
We condition on the measured $I_{\rm obs}$ in this expression; its median
signal-to-noise ratio is approximately 220, and its propagated uncertainty
is negligible relative to the uncertainties in $Q$ and $U$. The likelihood
is defined by
\begin{align}
\chi^2={}&
\Delta\boldsymbol Q^{T}C_Q^{-1}\Delta\boldsymbol Q
+\Delta\boldsymbol U^{T}C_U^{-1}\Delta\boldsymbol U,
\nonumber\\
\Delta\boldsymbol Q&\equiv\boldsymbol Q-\boldsymbol Q_{\rm mod},
\qquad
\Delta\boldsymbol U\equiv\boldsymbol U-\boldsymbol U_{\rm mod}.
\label{eq:J1025_chi2_spectra}
\end{align}
The covariance matrices include correlations between adaptive wavelength
bins; their mean correlation coefficient for adjacent bins is approximately
$0.09$. Cross-covariance between $Q$ and $U$ is taken to vanish, as expected
from their construction and confirmed by the data provider. Fitting $Q$ and
$U$ directly avoids a nonlinear transformation of their uncertainties and
the positive Ricean bias of $p$ in low-signal bins.

After excluding the telluric B band and selected line-contaminated regions,
we retain 69 of the supplied adaptive bins: 40 line-free continuum bins and
29 bins spanning the H$\alpha$ window. An inverse-variance-weighted linear fit
to the supplied $I$ values in the 40 continuum bins defines
$I_{\rm cont}(\lambda)$ and hence $f_{\rm cont}(\lambda)$. The likelihood
therefore contains 138 Stokes-flux measurements. At each $(\dot m,i)$
point in the precomputed thick-disk grid, we minimize over the three
continuous parameters $p_{\rm max}$, $\psi_{\rm disk}$, and
$\psi_{\rm dic}$. Counting also the two scanned physical parameters gives 133 nominal degrees of freedom. No continuum intercept, Stokes slope, continuum polarization, or continuum-subtracted H$\alpha$ summary statistic is included separately in the likelihood, because all of these quantities are derived from the same fitted spectrum. The best-fit model is shown in Figure~\ref{fig:pol_fit}, and the principal results are summarized in Table~\ref{tab:J1025_fit}. The formal minimum occurs at $\dot m=31.7$ and $i=70^\circ$, with $\chi^2=210.8$ and a global nominal $\chi^2/\mathrm{dof}=1.59$. However, the fit exhibits an almost complete degeneracy between accretion rate and inclination. A model with $\dot m=78.8$ and $i=65^\circ$ differs by only $\Delta\chi^2=0.2$, while the optimized values of $p_{\rm max}$, $\psi_{\rm disk}$, and $\psi_{\rm dic}$ are essentially unchanged. The polarization data therefore constrain primarily the continuum polarization amplitude rather than $\dot m$ and $i$ separately.

\begin{figure}[!ht]
\centering
\includegraphics[width=0.9\hsize,trim=0 0 0 0,clip]{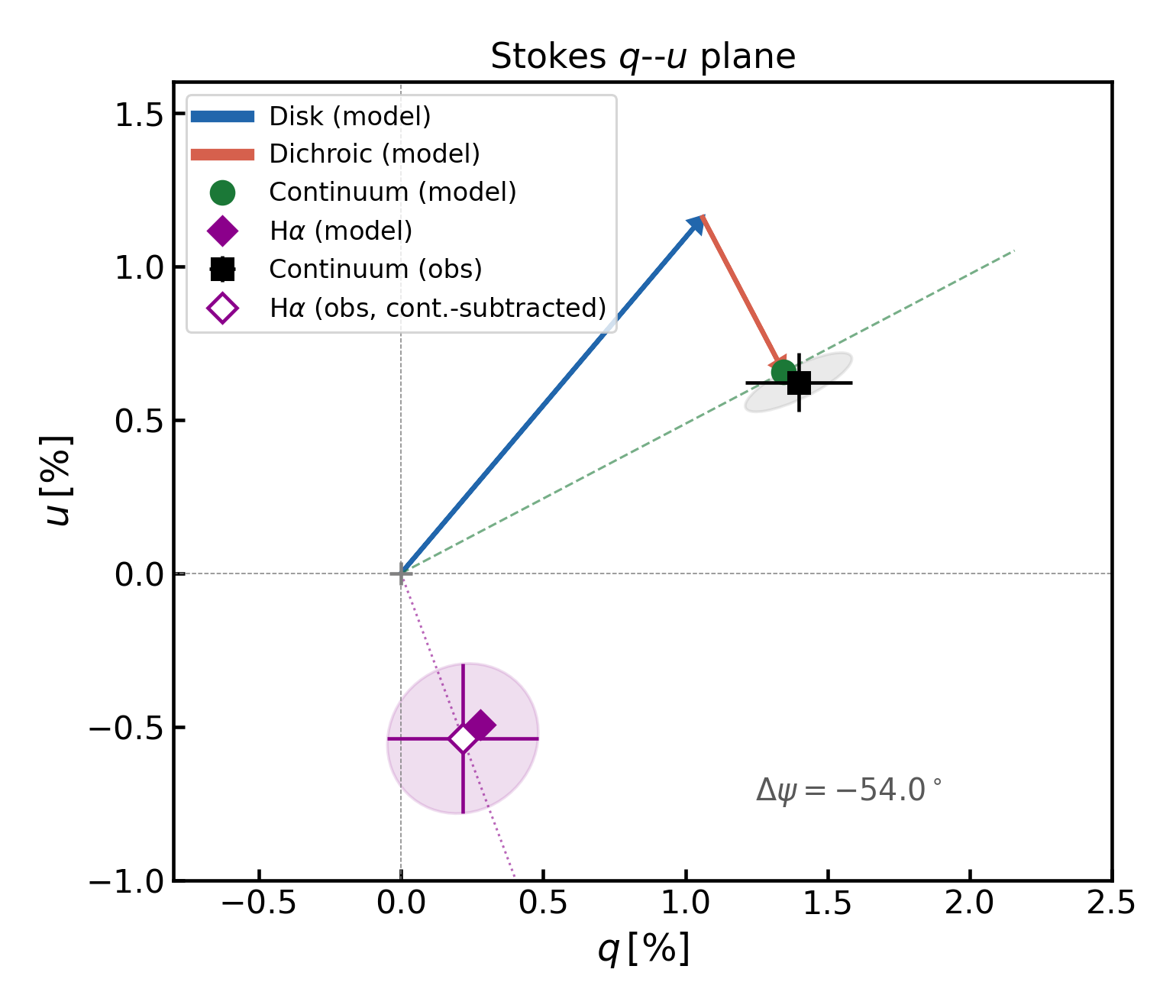}
\caption{Best-fit model in the Stokes $q$--$u$ plane. Blue and red arrows
show the intrinsic disk and dichroic vectors; their sum gives the continuum
prediction (green circle), compared with the measurement (black square).
The filled and open magenta diamonds show the model dichroic polarization at
H$\alpha$ and the published continuum-subtracted measurement, respectively;
the latter is diagnostic only. Ellipses and error bars show $1\sigma$
uncertainties, and the lines mark the corresponding position angles. The
component-angle separation is
$\Delta\psi\equiv\psi_{\rm dic}-\psi_{\rm disk}$.}
\label{fig:pol_qu}
\end{figure}

The decomposition of the best-fitting Stokes vectors is shown in
Figure~\ref{fig:pol_qu}, while Table~\ref{tab:J1025_fit} lists the fitted
parameters and derived polarization components. The disk and dichroic vectors
point in different directions and partially cancel. For plausible extinction
parameters, \(A_V\simeq2\)--4 mag and \(R_V\simeq3\)--4, the fitted dichroic
amplitude corresponds to
\(p_{\rm max}/E(B-V)\simeq0.4\)--\(1.2\%\,{\rm mag}^{-1}\), or
approximately \(5\)--\(13\%\) of the classical Galactic upper envelope.
This low effective efficiency need not imply weak grain alignment. Galactic
sightlines show that $p/A_V$ can decline above $A_V\sim1$ mag because
polarization from dust-bearing regions with different magnetic-field
orientations partially cancels \citep{Bijas2022}.

As a diagnostic, we fit the total and polarized-flux continua as
\(I_\lambda\propto\lambda^{\beta_I}\) and \(pI_\lambda\propto\lambda^{\beta_{p\times I}}\), respectively. The model predicts \(\Delta\beta\equiv\beta_{p\times I}-\beta_I=-0.03\), compared with the observed value of \(-0.80\pm0.57\), a difference of only \(1.4\sigma\). Direct linear fits to the continuum bins yield slopes of \(-0.19\pm0.16\%\) and \(-0.19\pm0.17\%\) per \(1000\,\)\AA\ in \(q\) and \(u\), respectively, whereas the corresponding model slopes are \(-0.011\%\) and \(+0.005\%\) per \(1000\,\)\AA. These differences are likewise only \(1.1\)--\(1.2\sigma\). The current data therefore do not reveal a statistically significant wavelength-dependent departure from the nearly gray model. Because \(\Delta\beta\) and the fitted continuum slopes are derived from the same spectral bins used in the fit, they are reported only as diagnostics and are not included as additional terms in Equation~(\ref{eq:J1025_chi2_spectra}).

The directly measured polarization-angle offset between the continuum and continuum-subtracted broad H$\alpha$ is
$\Delta\vartheta=-48^\circ\pm4^\circ\,({\rm stat})
\pm12^\circ\,({\rm epoch\ systematic})$ \citep{Deugenio2026_pol}. The model
offset is $-43^\circ$, consistent with the measurement once the
epoch-to-epoch systematic uncertainty is included. It arises because the
continuum angle is set by the vector sum of the intrinsic disk and dichroic
components, whereas the broad-line angle reflects the dichroic component
alone. The latter is set by the mean projected magnetic-field orientation in
the circumnuclear dusty screen, which need not coincide with the projected
disk axis. No geometric asymmetry of the accretion flow is therefore required
to explain the observed position-angle offset. The $\pm12^\circ$ term is an
empirical estimate of the angle variation among independently reduced
observing-epoch or observing-block groupings. It is distinct from both the
formal statistical uncertainty propagated through the Stokes covariance and
the instrumental polarization floor. No resolved radio jet, ionization cone,
or other independent projected-axis position angle is presently available
for J1025+1402, so the absolute disk-axis angle remains a nuisance parameter;
resolved radio or narrow-line imaging would provide a direct future test.

\begin{table*}
\caption{Direct full-spectrum polarization fit to SDSS~J1025+1402.}
\label{tab:J1025_fit}
\centering
\setlength{\tabcolsep}{10pt}
\begin{tabular}{lll}
\hline\hline
Quantity & Value & Notes \\
\hline
\multicolumn{3}{l}{\textit{Data and likelihood}} \\[2pt]
Retained bins & $69$ & $40$ continuum $+29$ H$\alpha$ \\
Fitted data values & $138$ & $Q_k$ and $U_k$ in every retained bin \\
$\chi^2$ & $210.8$ & continuum: $76$; H$\alpha$: $135$ \\
dof & $133$ & three continuous $+$ two grid parameters \\
nominal $\chi^2/{\rm dof}$ & $1.59$ & $210.8/133$ \\
\hline
\multicolumn{3}{l}{\textit{Best grid point and fitted parameters}} \\[2pt]
$\dot m$ & $31.7$ & grid minimum \\
$i$ & $70^\circ$ & grid minimum \\
$p_{\rm max}$ & $0.58\pm0.08\%$ & formal local error \\
$\psi_{\rm disk}$ & $24^\circ\pm2^\circ$ & formal local error \\
$\psi_{\rm dic}$ & $-30^\circ\pm3^\circ$ & formal local error \\[4pt]
\hline
\multicolumn{3}{l}{\textit{Derived diagnostics}} \\[2pt]
$p_{\rm disk}$ & $1.6\%$ & mean intrinsic disk polarization \\
$p_{\rm cont}$ & $1.5\%$ & mean continuum model \\
$\vartheta_{\rm cont}$ & $13^\circ$ & mean continuum model \\
$p_{{\rm dic},\,{\rm H}\alpha}$ & $0.6\%$ & line polarization in the model \\
$\psi_{\rm dic}-\psi_{\rm disk}$ & $-54^\circ$ & component-angle difference \\
$p_{\rm max}/E(B-V)$ & $0.4$--$1.2\%\,{\rm mag}^{-1}$ &
for $A_V=2$--4 mag and $R_V=3$--4 \\
$\Delta\beta$ & $-0.03$ & observed: $-0.80\pm0.57$; not fitted \\[4pt]
\hline
\end{tabular}
\tablefoot{The likelihood contains the directly measured $Q$ and $U$ fluxes
and their full wavelength covariance matrices, rather than continuum or
H$\alpha$ summary statistics. The quoted uncertainties on
the three continuous parameters are local-curvature estimates rescaled by
$\sqrt{\chi^2_\nu}$. The Serkowski parameters
$\lambda_{\rm max}=5500\,$\AA\ and $K=0.92$ are fixed.}
\end{table*}

\subsection{The H$\alpha$ core: a polarized narrow component?}
\label{sec:J1025_core}

The baseline fit leaves systematic residuals within a few hundred
$\mathrm{km\,s^{-1}}$ of the H$\alpha$ line center.
The spatially extended narrow-line region supplies most of the intensity at
the peak, so its polarization need not follow the nuclear continuum. We
therefore extend the model with a polarized narrow component and refit the same
69 bins.

We continue to neglect intrinsic or scattered BLR polarization. The continuum
carries the disk and dichroic components, while broad H$\alpha$ carries only
the dichroic component. Emission from narrow H$\alpha$ and [N II] is assumed to
originate outside the dichroic screen and to be intrinsically unpolarized. A
fraction of these photons may nevertheless acquire polarization by scattering
into the line of sight. Let \(f_{\rm NL}(\lambda)\) be the fraction of the
observed intensity contributed by the unscattered narrow-line emission.
The extended model is
\begin{align}
q_{\rm mod}(\lambda)
&=[1-f_{\rm NL}(\lambda)]q_{\rm dic}(\lambda)
 +f_{\rm cont}(\lambda)q_{\rm disk}(\lambda)\nonumber\\
&\quad +a_q\Phi(\lambda)f_{\rm cont}(\lambda),\nonumber\\[2pt]
u_{\rm mod}(\lambda)
&=[1-f_{\rm NL}(\lambda)]u_{\rm dic}(\lambda)
 +f_{\rm cont}(\lambda)u_{\rm disk}(\lambda)\nonumber\\
&\quad +a_u\Phi(\lambda)f_{\rm cont}(\lambda),
\label{eq:J1025_scatter}
\end{align}
Here $a_q$ and $a_u$ are polarized Stokes-flux amplitudes normalized to the
adjacent continuum intensity. Their position angle satisfies
$\tan 2\psi_{\rm sc}=a_u/a_q$, with the quadrant fixed by the signs of $a_q$
and $a_u$. The template $\Phi$ combines narrow H$\alpha$ and [N II], with
relative strengths tied to the total-intensity decomposition. Each seed line
is convolved with a symmetric two-sided exponential of velocity scale $v_e$;
the instrumental line-spread function and adaptive spectral bins are included
in the forward model. The intrinsic H$\alpha$ seed has
$\mathrm{FWHM}=92\,\mathrm{km\,s^{-1}}$ \citep{Ji2026}.

Refitting $p_{\rm max}$, $\psi_{\rm disk}$, and $\psi_{\rm dic}$ jointly with
$a_q$, $a_u$, and $v_e$ at the same grid point ($\dot m=31.7$, $i=70^\circ$)
gives the values in Table~\ref{tab:J1025_core}. The chi-square decreases by
$28.5$, from $210.8$ to $182.3$, for three additional parameters;
$\Delta\mathrm{AIC}=-22.5$ and $\Delta\mathrm{BIC}=-13.7$ for $n=138$ data
values, where AIC and BIC denote the Akaike and Bayesian information
criteria, respectively. An otherwise identical fit with unpolarized direct
narrow lines but no scattered component gives $\chi^2=239.1$; adding
scattering lowers this by $56.8$. These criteria favor the narrow-scattering
component. The continuum
parameters shift only modestly.
For gas at $10^4\,$K, the characteristic electron thermal speed is
$\simeq550\,\mathrm{km\,s^{-1}}$, about six times $v_e$, so thermal electron
motions would smear the scattered profile too broadly. Grain thermal speeds
are negligible, and the much larger optical opacity of dust in normally dusty
gas further favors dust scattering \citep{Capetti2021}.
Dust is inferred to be present in at least part of the NLR in many AGNs
\citep{NetzerLaor1993,Wills1993}, providing a plausible scattering medium.
The grains need not lie outside the NLR as a whole, but may be mixed with or
surround the compact inner narrow-line-emitting region.
The fitted $\psi_{\rm sc}=-44^\circ$ nominally differs from both
$\psi_{\rm disk}$ and $\psi_{\rm dic}$, suggesting an additional polarizing
structure, plausibly a polar scattering cone \citep{Antonucci1985}. The
published continuum-subtracted broad-H$\alpha$ angle,
$-34^\circ\pm13^\circ$, lies between
$\psi_{\rm sc}$ and $\psi_{\rm dic}$, as expected for a blend of scattered
narrow-line and dichroically polarized broad-line flux.

\begin{figure}
\centering
\includegraphics[width=0.9\hsize]{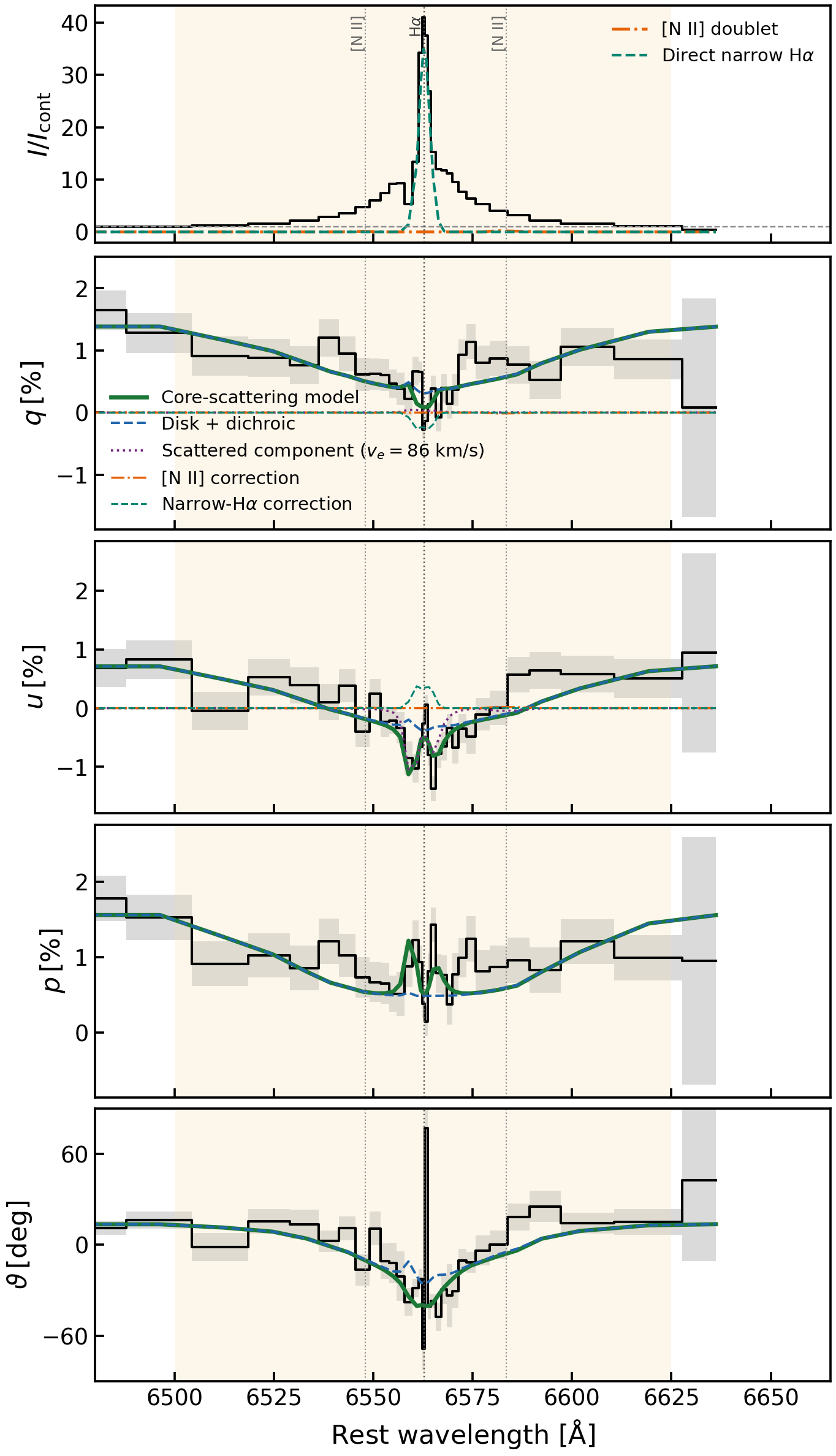}
\caption{Polarized narrow-component fit to the H$\alpha$+[N II] core.
The fitted scattered component has
$v_e=86\,\mathrm{km\,s^{-1}}$ and $\psi_{\rm sc}=-43.8^\circ$.
The top panel shows the continuum-normalized intensity and direct narrow-line
components. Lower panels show $q$, $u$, $p$, and the polarization position
angle $\vartheta$: black histograms and gray bands are the data and $1\sigma$
uncertainties, green is the extended model, and dashed blue is the
disk-plus-dichroic baseline. Teal and orange curves in the $q$ and $u$ panels
show the removal of dichroic polarization over the direct narrow-line flux;
separate component amplitudes are not shown in the nonlinear $p$ and
$\vartheta$ panels. The position angle is poorly constrained where $p$ is
small.
Shading marks the fitted window and dotted lines the three rest wavelengths.
The displayed Stokes profiles are $q=Q/I_{\rm obs}$ and
$u=U/I_{\rm obs}$; the likelihood is evaluated directly in $Q$ and $U$ flux.
The extended model has nominal
$\chi^2/{\rm dof}=182.3/130=1.40$.}
\label{fig:pol_core}
\end{figure}

\begin{table*}
\caption{Polarized narrow-component fit to the H$\alpha$+[N II] core.}
\label{tab:J1025_core}
\centering
\setlength{\tabcolsep}{10pt}
\begin{tabular}{lll}
\hline\hline
Quantity & Value & Notes \\
\hline
\multicolumn{3}{l}{\textit{Model setup}} \\[2pt]
$\dot m$ & $31.7$ & inherited from Table~\ref{tab:J1025_fit} \\
$i$ & $70^\circ$ & inherited from Table~\ref{tab:J1025_fit} \\
Seed FWHM & $92\,{\rm km\,s^{-1}}$ & fixed\tablefootmark{a} \\[4pt]
\hline
\multicolumn{3}{l}{\textit{Continuum: intrinsic disk polarization}} \\[2pt]
$\psi_{\rm disk}$ & $23^\circ\pm3^\circ$ & formal local error \\[4pt]
\hline
\multicolumn{3}{l}{\textit{Dichroic screen: acts on continuum and broad H$\alpha$}} \\[2pt]
$p_{\rm max}$ & $0.5\pm0.1\%$ & formal local error \\
$\psi_{\rm dic}$ & $-27^\circ\pm3^\circ$ & formal local error \\
$\psi_{\rm dic}-\psi_{\rm disk}$ & $-50^\circ$ & component-angle difference \\[4pt]
\hline
\multicolumn{3}{l}{\textit{Narrow H$\alpha$ and [N II]: direct component}} \\[2pt]
Direct narrow H$\alpha$ & unpolarized & $93\%$ of peak intensity \\
Direct [N II] & unpolarized & $6\%$ of peak intensity \\[4pt]
\hline
\multicolumn{3}{l}{\textit{Narrow H$\alpha$ and [N II]: scattered component}} \\[2pt]
$a_q$ & $1\pm3\%$ & relative to adjacent continuum intensity \\
$a_u$ & $-19\pm4\%$ & relative to adjacent continuum intensity \\
$\psi_{\rm sc}$ & $-44^\circ\pm4^\circ$ & derived formal error \\
$f_{\rm sc}p_{\rm sc}({\rm H}\alpha)$ & $0.9\pm0.2\%$ &
integrated polarized/direct narrow-line flux; derived \\
$v_e$ & $86\pm30\,{\rm km\,s^{-1}}$ & formal local error \\
$\psi_{\rm sc}-\psi_{\rm dic}$ & $-16^\circ\pm6^\circ$ & derived formal error \\[4pt]
\hline
\multicolumn{3}{l}{\textit{Goodness of fit and model comparison}} \\[2pt]
$\chi^2$ & $182.3$ & baseline on same bins: $210.8$ \\
nominal $\chi^2/{\rm dof}$ & $1.40$ & $182.3/130$; eight model parameters \\
$\Delta\chi^2$ & $-28.5$ & for three added parameters \\
$\Delta{\rm AIC}$ & $-22.5$ & relative to Table~\ref{tab:J1025_fit} \\
$\Delta{\rm BIC}$ & $-13.7$ & relative to Table~\ref{tab:J1025_fit} \\
No-scatter $\chi^2$ & $239.1$ & same unpolarized direct narrow lines \\
$\chi^2$ reduction vs. no scatter & $56.8$ & relative to the no-scatter fit \\[4pt]
\hline
\multicolumn{3}{l}{\textit{Not fitted}} \\[2pt]
$\vartheta_{{\rm H}\alpha}$ & $-34^\circ\pm13^\circ$ &
continuum-subtracted broad line\tablefootmark{b} \\[4pt]
\hline
\end{tabular}
\tablefoot{The fit uses the same data and likelihood as
Table~\ref{tab:J1025_fit}. The disk grid point is fixed, while the three
continuum and three scattering parameters are refitted jointly. The quoted
uncertainties on fitted parameters are local-curvature estimates rescaled by
$\sqrt{\chi^2_\nu}$; uncertainties
on derived quantities include correlations among the fitted parameters.
\tablefoottext{a}{Seed width from \citet{Ji2026}.}
\tablefoottext{b}{Published value from \citet{Deugenio2026_pol}; the quoted
uncertainty includes the epoch systematic.}}
\end{table*}

The fitted quantity $f_{\rm sc}p_{\rm sc}({\rm H}\alpha)$ is the integrated
scattered polarized H$\alpha$ flux relative to the direct, unscattered narrow
H$\alpha$ flux. Because $a_q$ and $a_u$ multiply the same H$\alpha$-only
template $\Phi_{{\rm H}\alpha}$ at the single angle $\psi_{\rm sc}$, it is
given by
\begin{multline}
f_{\rm sc}p_{\rm sc}({\rm H}\alpha)
\equiv
\sqrt{a_q^2+a_u^2}\,
\\
\frac{\int \Phi_{{\rm H}\alpha}(\lambda)I_{\rm cont}(\lambda)
\,\mathrm{d}\lambda}
{\int I_{{\rm H}\alpha,{\rm direct}}(\lambda)\,\mathrm{d}\lambda}
=0.9\pm0.2\%.
\end{multline}
The fit
also treats Balmer absorption as unpolarized and fixes the seed width from the
total-intensity decomposition, because the polarimetric data cannot constrain
it independently.

\section{Discussion and Summary}

Azimuthal cancellation suppresses the disk-integrated polarization at low
inclination, as it does for any axisymmetric disk atmosphere. The distinctive
super-Eddington behavior arises from the curved photosphere, self-shadowing,
and wavelength-dependent dilution by the outer thin disk: together these
effects allow the polarization to reach several percent toward edge-on views
and control its UV-to-optical color. The polarization also increases with
$\dot m$, because the unpolarized outer thin disk contributes less dilution
as $r_{\rm thick}$ expands. High-$\dot m$ models are consequently nearly
gray, whereas lower-rate models show a larger UV-to-optical decline.

We have also tested the sensitivity to weak polarization of
the outer thin disk. At fixed $(\dot m,i)=(31.7,70^\circ)$, the thin disk
supplies $17\%$ of the total disk flux at $1500\,$\AA\ and $27$--$28\%$ over
$5300$--$6500\,$\AA. Relative to the unpolarized-thin-disk baseline, assigning
it a wavelength-independent polarization of $0.5\%$ shifts the net intrinsic
disk polarization by $\pm0.09$ percentage points at $1500\,$\AA\ and
$\pm0.13$--$0.14$ percentage points in the optical;
for $1\%$ polarization these shifts double to $\pm0.17$ and
$\pm0.27$--$0.28$ percentage points. The positive sign applies when the thin-
and thick-disk polarization directions are parallel, and the negative sign
when they are perpendicular. For a coplanar, plane-parallel electron-scattering
atmosphere, the outer thin-disk polarization is parallel to the integrated
thick-flow polarization in this best-fitting geometry; the perpendicular case
is retained only as a conservative limiting case for a possible reversal of
the unresolved atmospheric Stokes sign. The parallel cases reduce the UV-to-optical
variation: $p_{\rm disk}(1500\,\text{\AA})/p_{\rm disk}(6500\,\text{\AA})$
decreases from $1.06$ in the unpolarized-thin-disk calculation to $1.03$ and
$1.00$ for thin-disk polarizations of $0.5\%$ and $1\%$, respectively. The
perpendicular cases instead increase this ratio to $1.10$ and $1.15$. The
baseline ratio is smaller than the value expected from thin-disk dilution
alone because wavelength-dependent radial weighting makes the polarization of
the thick component itself approximately $8\%$ larger at $6500\,$\AA\ than
at $1500\,$\AA. Across
the much narrower observed $5300$--$6650\,$\AA\ interval, the thin-disk flux
fraction changes very little, so the main effect is a shift in polarization
amplitude rather than an additional measurable slope. In a full refit, a
parallel thin-disk contribution would tend to require either stronger
dichroic cancellation or a smaller thick-flow polarization, whereas a
perpendicular contribution would favor weaker dichroic cancellation or a
larger thick-flow contribution. This systematic therefore broadens the
existing $\dot m$--$i$ and component-amplitude degeneracies but does not alter
the qualitative interpretation of the continuum and broad-line Stokes
vectors.

Applied to the spectropolarimetric data of SDSS~J1025+1402, the model favors a
high-accretion-rate, high-inclination geometry but leaves a strong
$\dot m$--$i$ degeneracy. The observed continuum is described by the vector
sum of intrinsic disk polarization and dichroic polarization from the
circumnuclear dusty component invoked in the LRD--LBD unification model.
Broad H$\alpha$ is assumed to be intrinsically unpolarized and to pass through
the same screen, so it retains only the dichroic component, whose position
angle is set by the mean projected magnetic-field orientation in the dusty
screen. The continuum-to-line offset therefore reflects the different
orientations of the disk and dichroic Stokes vectors and does not require a
non-axisymmetric disk. This transmission geometry is also consistent with the
absence of a position-angle swing across broad H$\alpha$, because the screen
applies one orientation to all line velocities.

Residuals at the H$\alpha$+[N II] core motivate an additional component. If the direct narrow lines originate outside the dichroic screen and are unpolarized, allowing some of their photons to scatter into the line of sight decreases $\chi^2$ by 28 for three added parameters relative to the baseline
($\Delta\mathrm{BIC}=-13$), and by 57 relative to the corresponding unpolarized-narrow, no-scatter fit. The fitted scattering-polarization angle suggests a polar structure, while the narrow velocity scale favors dust over thermal electrons. This interpretation remains provisional: it fixes the seed-line decomposition and treats Balmer absorption as unpolarized.

The distinction between intrinsic disk polarization and an external scattering origin is relevant to comparisons with ordinary quasars. In the VLT/FORS2 sample of \citet{Capetti2021}, the continuum and integrated broad H$\alpha$ generally have similar position angles, and swings across the line
are common. Those properties are naturally produced when one equatorial scatterer redirects both continuum and BLR photons. J1025+1402 instead shows a large continuum-to-line offset and no broad-line swing \citep{Deugenio2026_pol}. A single equatorial scatterer does not naturally produce both features, whereas a disk-polarized continuum viewed through a dichroic screen does. The modest optical polarization of most radio-quiet
Narrow-Line Seyfert~1 galaxies may likewise reflect their predominantly lower inclinations; individual objects nevertheless show blue-rising polarization and line-dependent Stokes structure consistent with orientation playing an important role \citep{Goodrich1989,Leighly1997,Smith2004}.
Appendices~\ref{app:equatorial_scattering} and
\ref{app:equatorial_dichroic} test this standard alternative alone and with a
foreground dichroic screen. The screen partly cancels the broad-line signal,
but the best combined fit remains substantially worse than the
disk-plus-dichroic model.

The polarization of J1025+1402 does not require an external
continuum-scattering component: its observed amplitude is reproduced by the disk-plus-dichroic model. More generally, 
continuum polarization substantially exceeding the few-percent levels predicted by the disk-plus-dichroic model could signal an additional contribution from polar scattering. Continuum radiation preferentially emitted at low inclination could be scattered into a high-inclination line of sight by polar material, as in the classical interpretation of Seyfert~2 polarization \citep{Antonucci1985,Smith2004}. Unlike in a Seyfert~2 nucleus, the attenuated direct equatorial continuum would remain visible and dilute the scattered component. In our SED models, the low-inclination optical continuum is brighter than the edge-on continuum by a factor of approximately 5--10
\citep{Madau2026}. Redirecting an effective $10\%$ of this radiation toward the observer could therefore make the scattered flux comparable to the direct equatorial flux and, for an angle-averaged scattered-light polarization of approximately $30\%$, produce an observed polarization of approximately
$10$--$15\%$, neglecting host-galaxy dilution and vector cancellation. Such a high polarization of the scattered light is possible for optically thin scattering near right angles, but depends on the opening angle, optical depth, and scattering phase function. An unattenuated scattered component this strong
would be too blue to preserve the red optical colors of an LRD and would therefore also have to suffer substantial extinction along its path to the observer, although not necessarily through the same dust screen as the direct continuum. The argument applies to the anisotropic continuum and does not apply to the spatially extended, isotropically emitted narrow-line radiation.

We neglect general-relativistic propagation of the optical/UV Stokes vector.
Appendix~\ref{app:gr_origin} shows that this is adequate for the
high-inclination solutions relevant here: after all orders of internal
wall reflection are included, photons born at cylindrical radii
$r<20r_s$ supply less than $1.1\%$ of the received flux, while self-shadowing
ensures that none of the observed $|\mathrm dQ|$ is generated at a final
polarizing surface inside $20r_s$ at $i=65^\circ$--$70^\circ$.

We also neglect magneto-optical transfer in the geometrically
thick flow. Faraday rotation has little observational consequence for the
inner-funnel signal where it is already suppressed by azimuthal cancellation
at low inclination or by self-obscuration at the high inclinations relevant
here. Any appreciable effect on the fitted signal would instead have to rotate
or depolarize the radiation emerging from the outer rim of the thick flow,
which dominates the observed polarized flux. Our radiation-supported solution
does not specify the photospheric magnetic-field strength or topology, so this
correction cannot presently be quantified. The polarization calculated here
should therefore be regarded as the unmagnetized limit; substantial Faraday
depolarization of the outer rim would lower the intrinsic polarization and
broaden the inferred $\dot m$--$i$ degeneracy.

Two limitations are central. First, the Chandrasekhar law is an idealized closure for a semi-infinite, plane-parallel, conservative atmosphere. The funnel wall is curved, has spatially varying optical depth, and receives anisotropic irradiation. Our scalar Lambertian wall-to-wall coupling followed by an angle-dependent polarized escape calculation therefore yields approximate polarization fractions. The full-Stokes Monte Carlo benchmark in Appendix~\ref{app:stokes_benchmark} agrees with the approximation within $0.6\sigma$ in all four low-inclination bins, where the funnel is directly visible. At the high inclinations relevant to the fit, the inner funnel is self-obscured and its polarized contribution becomes negligible; the observed signal is instead dominated by the outer rim of the thick flow. Second, the present optical baseline is too short to separate $\dot m$ from $i$ or constrain the model's weak color dependence.

A broader wavelength range offers further tests. For the best-fitting
$(\dot m,i)=(31.7,70^\circ)$ model, the predicted total polarization is
approximately $1.63\%$, $1.50\%$, $1.49\%$, and $1.48\%$ at $1500\,$\AA,
$5500\,$\AA, $6500\,$\AA, and $1\,\mu$m, respectively. The corresponding
position angles are approximately $22^\circ$, $13^\circ$, $13^\circ$, and
$16^\circ$. These extrapolations are model-dependent because grain size,
composition, alignment, and magnetic-field disorder can change the dichroic
wavelength dependence away from the adopted Serkowski curve
\citep{Whittet1992,Andersson2015}. UV or blue spectropolarimetry can therefore
probe the relative roles of nearly gray electron scattering and
wavelength-dependent dichroic polarization, while near-infrared data can
trace the declining dichroic tail. UV emission from young stars could also
dilute the AGN polarization or contribute a separate Stokes vector if it is
itself dichroically polarized. The orientation-based population model further
predicts that LRDs should, on average, be more polarized than unobscured LBDs,
with the largest rest-frame UV polarization in the most inclined and rapidly
accreting sources \citep{MadauMaiolino2026,Madau_LF2026}. Conversely,
low-inclination objects should remain weakly polarized because of
cancellation.

In summary, the continuum polarization of J1025+1402 is consistent with a geometrically thick, rapidly accreting flow seen at high inclination and combined with modest dichroic transmission. The narrow-line residuals provide
tentative evidence for a separate dusty scattering region. Spectropolarimetry therefore supplies a geometric test of the orientation-based little-dot model that is complementary to SED fitting and emission-line diagnostics.

\begin{acknowledgements}
We thank Ari Laor for useful discussions on accretion-disk polarization.
\end{acknowledgements}

\label{lastpage}
\bibliographystyle{aa}
\bibliography{paper}

@ARTICLE{Capetti2021,
       author = {{Capetti}, Alessandro and {Laor}, Ari and {Baldi}, Ranieri D. and {Robinson}, Andrew and {Marconi}, Alessandro},
        title = "{Spectropolarimetry of low redshift quasars: origin of the polarization and implications for black hole mass estimates}",
      journal = {\mnras},
         year = 2021,
        month = apr,
       volume = {502},
       number = {4},
        pages = {5086-5103},
          doi = {10.1093/mnras/stab279},
archivePrefix = {arXiv},
       eprint = {2102.05935},
 primaryClass = {astro-ph.GA},
       adsurl = {https://ui.adsabs.harvard.edu/abs/2021MNRAS.502.5086C}
}

@ARTICLE{NetzerLaor1993,
       author = {{Netzer}, Hagai and {Laor}, Ari},
        title = "{Dust in the Narrow-Line Region of Active Galactic Nuclei}",
      journal = {\apjl},
         year = 1993,
        month = feb,
       volume = {404},
        pages = {L51-L54},
          doi = {10.1086/186741},
       adsurl = {https://ui.adsabs.harvard.edu/abs/1993ApJ...404L..51N}
}

@ARTICLE{Wills1993,
       author = {{Wills}, Beverley J. and {Netzer}, Hagai and {Brotherton}, Michael S. and {Han}, Mingsheng and {Wills}, D. and {Baldwin}, Jack A. and {Ferland}, Gary J. and {Browne}, I.~W.~A.},
        title = "{The Narrow-Line Region of High-Luminosity Active Galactic Nuclei}",
      journal = {\apj},
         year = 1993,
        month = jun,
       volume = {410},
        pages = {534-542},
          doi = {10.1086/172772},
       adsurl = {https://ui.adsabs.harvard.edu/abs/1993ApJ...410..534W}
}

@ARTICLE{Smith2005,
       author = {{Smith}, J.~E. and {Robinson}, A. and {Young}, S. and {Axon}, D.~J. and {Corbett}, Elizabeth A.},
        title = "{Equatorial scattering and the structure of the broad-line region in Seyfert nuclei: evidence for a rotating disc}",
      journal = {\mnras},
         year = 2005,
        month = may,
       volume = {359},
       number = {3},
        pages = {846-864},
          doi = {10.1111/j.1365-2966.2005.08895.x},
archivePrefix = {arXiv},
       eprint = {astro-ph/0501640},
 primaryClass = {astro-ph},
       adsurl = {https://ui.adsabs.harvard.edu/abs/2005MNRAS.359..846S}
}

@ARTICLE{Lira2020,
       author = {{Lira}, P. and {Goosmann}, R.~W. and {Kishimoto}, M. and {Cartier}, R.},
        title = "{The polarized signal from broad emission lines in AGNs}",
      journal = {\mnras},
         year = 2020,
        month = jan,
       volume = {491},
       number = {1},
        pages = {1-12},
          doi = {10.1093/mnras/stz2774},
archivePrefix = {arXiv},
       eprint = {1906.08718},
 primaryClass = {astro-ph.GA}
}

@ARTICLE{Antonucci1985,
       author = {{Antonucci}, R.~R.~J. and {Miller}, J.~S.},
        title = "{Spectropolarimetry and the nature of NGC 1068.}",
      journal = {\apj},
         year = 1985,
        month = oct,
       volume = {297},
        pages = {621-632},
          doi = {10.1086/163559},
       adsurl = {https://ui.adsabs.harvard.edu/abs/1985ApJ...297..621A}
}

@ARTICLE{Deugenio2026_pol,
       author = {{D'Eugenio}, Francesco and {Pezzulli}, Gabriele and {Maiolino}, Roberto and {Marconi}, Alessandro and {Ji}, Xihan and {Ramos Almeida}, Cristina and {Ferrara}, Andrea and {Madau}, Piero and {Lin}, Xiaojing and {Acosta-Pulido}, Jos{\'e} A. and {Bian}, Fuyan and {Brazzini}, Matilde and {Cai}, Zheng and {Carniani}, Stefano and {Fan}, Xiaohui and {Juod{\v{z}}balis}, Ignas and {Pascalau}, Robert G. and {Scholtz}, Jan and {Simmonds}, Charlotte and {Sun}, Fengwu and {Tacchella}, Sandro},
        title = "{Misaligned or chaotic? A strong break of axial symmetry in the local LRD J1025 revealed with VLT/FORS2 spectropolarimetry}",
      journal = {arXiv e-prints},
         year = 2026,
        month = jul,
          eid = {arXiv:2607.28715},
        pages = {arXiv:2607.28715},
archivePrefix = {arXiv},
       eprint = {2607.28715},
 primaryClass = {astro-ph.GA},
       adsurl = {https://ui.adsabs.harvard.edu/abs/2026arXiv260728715D}
}

@ARTICLE{Smith2002,
       author = {{Smith}, J.~E. and {Young}, S. and {Robinson}, A. and {Corbett}, E.~A. and {Giannuzzo}, M.~E. and {Axon}, D.~J. and {Hough}, J.~H.},
        title = "{A spectropolarimetric atlas of Seyfert 1 galaxies}",
      journal = {\mnras},
         year = 2002,
        month = sep,
       volume = {335},
       number = {3},
        pages = {773-798},
          doi = {10.1046/j.1365-8711.2002.05665.x},
archivePrefix = {arXiv},
       eprint = {astro-ph/0205204},
 primaryClass = {astro-ph},
       adsurl = {https://ui.adsabs.harvard.edu/abs/2002MNRAS.335..773S}
}

@ARTICLE{Ji2026,
       author = {{Ji}, Xihan and {D'Eugenio}, Francesco and {Juod{\v{z}}balis}, Ignas and {Walton}, Dominic J. and {Fabian}, Andrew C. and {Maiolino}, Roberto and {Ramos Almeida}, Cristina and {Acosta Pulido}, Jose A. and {Belokurov}, Vasily A. and {Isobe}, Yuki and {Jones}, Gareth and {Maraston}, Claudia and {Scholtz}, Jan and {Simmonds}, Charlotte and {Tacchella}, Sandro and {Terlevich}, Elena and {Terlevich}, Roberto},
        title = "{Lord of LRDs: insights into a 'Little Red Dot' with a low-ionization spectrum at z = 0.1}",
      journal = {\mnras},
         year = 2026,
        month = jan,
       volume = {545},
       number = {3},
          eid = {staf2235},
        pages = {staf2235},
          doi = {10.1093/mnras/staf2235},
archivePrefix = {arXiv},
       eprint = {2507.23774},
 primaryClass = {astro-ph.GA},
       adsurl = {https://ui.adsabs.harvard.edu/abs/2026MNRAS.545f2235J}
}

@ARTICLE{Serkowski1975,
       author = {{Serkowski}, K. and {Mathewson}, D.~S. and {Ford}, V.~L.},
        title = "{Wavelength dependence of interstellar polarization and ratio of total to selective extinction.}",
      journal = {\apj},
         year = 1975,
        month = feb,
       volume = {196},
        pages = {261-290},
          doi = {10.1086/153410},
       adsurl = {https://ui.adsabs.harvard.edu/abs/1975ApJ...196..261S}
}

@ARTICLE{Koratkar1999,
       author = {{Koratkar}, Anuradha and {Blaes}, Omer},
        title = "{The Ultraviolet and Optical Continuum Emission in Active Galactic Nuclei: The Status of Accretion Disks}",
      journal = {\pasp},
         year = 1999,
        month = jan,
       volume = {111},
       number = {755},
        pages = {1-30},
          doi = {10.1086/316294},
       adsurl = {https://ui.adsabs.harvard.edu/abs/1999PASP..111....1K}
}

@ARTICLE{Taverna2021,
       author = {{Taverna}, R. and {Marra}, L. and {Bianchi}, S. and {Dov{\v{c}}iak}, M. and {Goosmann}, R. and {Marin}, F. and {Matt}, G. and {Zhang}, W.},
        title = "{Spectral and polarization properties of black hole accretion disc emission: including absorption effects}",
      journal = {\mnras},
         year = 2021,
        month = mar,
       volume = {501},
       number = {3},
        pages = {3393-3405},
          doi = {10.1093/mnras/staa3859},
archivePrefix = {arXiv},
       eprint = {2012.06504},
 primaryClass = {astro-ph.HE},
       adsurl = {https://ui.adsabs.harvard.edu/abs/2021MNRAS.501.3393T}
}

@ARTICLE{Whittet1992,
       author = {{Whittet}, D.~C.~B. and {Martin}, P.~G. and {Hough}, J.~H. and {Rouse}, M.~F. and {Bailey}, J.~A. and {Axon}, D.~J.},
        title = "{Systematic Variations in the Wavelength Dependence of Interstellar Linear Polarization}",
      journal = {\apj},
         year = 1992,
        month = feb,
       volume = {386},
        pages = {562},
          doi = {10.1086/171039},
       adsurl = {https://ui.adsabs.harvard.edu/abs/1992ApJ...386..562W}
}

@ARTICLE{Sok2026,
       author = {{Sok}, Visal and {Nelson}, Erica J. and {Begelman}, Mitchell C. and {Dexter}, Jason and {D'Eugenio}, Francesco and {Greene}, Jenny E. and {Leja}, Joel and {Whitaker}, Katherine E. and {Bunker}, Andrew J. and {P{\'e}rez-Gonz{\'a}lez}, Pablo G. and {Rinaldi}, Pierluigi and {Torralba}, Alberto and {{\"U}bler}, Hannah},
        title = "{Constraints on the Gas Geometry Surrounding Little Red Dots through Narrow-Line Diagnostics}",
      journal = {arXiv e-prints},
         year = 2026,
        month = jun,
          eid = {arXiv:2606.23778},
        pages = {arXiv:2606.23778},
          doi = {10.48550/arXiv.2606.23778},
archivePrefix = {arXiv},
       eprint = {2606.23778},
 primaryClass = {astro-ph.GA},
       adsurl = {https://ui.adsabs.harvard.edu/abs/2026arXiv260623778S}
}

@ARTICLE{Smith2004,
       author = {{Smith}, J.~E. and {Robinson}, A. and {Alexander}, D.~M. and {Young}, S. and {Axon}, D.~J. and {Corbett}, Elizabeth A.},
        title = "{Seyferts on the edge: polar scattering and orientation-dependent polarization in Seyfert 1 nuclei}",
      journal = {\mnras},
         year = 2004,
        month = may,
       volume = {350},
       number = {1},
        pages = {140-160},
          doi = {10.1111/j.1365-2966.2004.07610.x},
archivePrefix = {arXiv},
       eprint = {astro-ph/0401496},
 primaryClass = {astro-ph},
       adsurl = {https://ui.adsabs.harvard.edu/abs/2004MNRAS.350..140S}
}

@ARTICLE{Leighly1997,
       author = {{Leighly}, Karen M. and {Kay}, Laura E. and {Wills}, Beverley J. and {Wills}, D. and {Grupe}, Dirk},
        title = "{The Optical Polarization and Warm Absorber in IRAS 17020+4544}",
      journal = {\apjl},
         year = 1997,
        month = nov,
       volume = {489},
        pages = {L137},
          doi = {10.1086/316793},
archivePrefix = {arXiv},
       eprint = {astro-ph/9709097},
 primaryClass = {astro-ph},
       adsurl = {https://ui.adsabs.harvard.edu/abs/1997ApJ...489L.137L}
}

@ARTICLE{Geris2026,
       author = {{Geris}, Sophia and {Maiolino}, Roberto and {Ji}, Xihan and {Risaliti}, Guido and {Lanzuisi}, Giorgio and {D'Eugenio}, Francesco and {Isobe}, Yuki and {Jones}, Gareth and {Harshan}, Anishya and {Brazzini}, Matilde and {Juod{\v{z}}balis}, Ignas and {Scholtz}, Jan and {Rinaldi}, Pierluigi and {{\"U}bler}, Hannah and {Baker}, William and {Bunker}, Andrew J. and {Brusa}, Marcella and {Carniani}, Stefano and {Charlot}, St{\'e}phane and {Curti}, Mirko and {Comastri}, Andrea and {Curtis-Lake}, Emma and {Gilli}, Roberto and {Hainline}, Kevin and {Madau}, Piero and {Marchesi}, Stefano and {Mazzolari}, Giovanni and {Napolitano}, Lorenzo and {Parlanti}, Eleonora and {Pentericci}, Laura and {Ramos Almeida}, Cristina and {Robertson}, Brant and {Silcock}, Maddie S. and {Tripodi}, Roberta and {Venturi}, Giacomo and {Vignali}, Cristian and {Vito}, Fabio and {Zhu}, Yongda},
        title = "{Little Red and Blue Dots: AGN-excited narrow lines, Lyman-$\alpha$ emission, and resemblance to standard quasars}",
      journal = {arXiv e-prints},
         year = 2026,
        month = jun,
          eid = {arXiv:2606.21614},
        pages = {arXiv:2606.21614},
          doi = {10.48550/arXiv.2606.21614},
archivePrefix = {arXiv},
       eprint = {2606.21614},
 primaryClass = {astro-ph.GA},
       adsurl = {https://ui.adsabs.harvard.edu/abs/2026arXiv260621614G}
}

@ARTICLE{Goodrich1989,
       author = {{Goodrich}, Robert W.},
        title = "{Spectropolarimetry of ``Narrow-Line'' Seyfert 1 Galaxies}",
      journal = {\apj},
         year = 1989,
        month = jul,
       volume = {342},
        pages = {224},
          doi = {10.1086/167586},
       adsurl = {https://ui.adsabs.harvard.edu/abs/1989ApJ...342..224G}
}

@BOOK{Chandrasekhar1960,
       author = {{Chandrasekhar}, Subrahmanyan},
        title = "{Radiative transfer}",
    publisher = {Dover Publications},
      address = {New York},
         year = 1960,
       adsurl = {https://ui.adsabs.harvard.edu/abs/1960ratr.book.....C}
}

@ARTICLE{Bijas2022,
       author = {{Bijas}, N. and {Eswaraiah}, Chakali and {Wang}, Jia-Wei and {Jose}, Jessy and {Chen}, Wen-Ping and {Li}, Di and {Lai}, Shih-Ping and {Ojha}, D.~K.},
        title = "{Revealing the dust grain polarization properties as a function of extinction and distance towards NGC 1893}",
      journal = {\mnras},
         year = 2022,
        month = sep,
       volume = {515},
       number = {3},
        pages = {3352-3369},
          doi = {10.1093/mnras/stac1927},
archivePrefix = {arXiv},
       eprint = {2207.03173},
 primaryClass = {astro-ph.GA},
       adsurl = {https://ui.adsabs.harvard.edu/abs/2022MNRAS.515.3352B}
}

@ARTICLE{Agol1998,
       author = {{Agol}, Eric and {Blaes}, Omer and {Ionescu-Zanetti}, Cristian},
        title = "{Polarization from magnetized accretion discs - II. The effects of absorption opacity on Faraday rotation}",
      journal = {\mnras},
         year = 1998,
        month = jan,
       volume = {293},
       number = {1},
        pages = {1-17},
          doi = {10.1046/j.1365-8711.1998.01107.x},
archivePrefix = {arXiv},
       eprint = {astro-ph/9612171},
 primaryClass = {astro-ph},
       adsurl = {https://ui.adsabs.harvard.edu/abs/1998MNRAS.293....1A}
}

@ARTICLE{Laor1990,
       author = {{Laor}, Ari and {Netzer}, Hagai and {Piran}, Tsvi},
        title = "{Massive thin accretion discs. II - Polarization}",
      journal = {\mnras},
         year = 1990,
        month = feb,
       volume = {242},
        pages = {560-569},
          doi = {10.1093/mnras/242.4.560},
       adsurl = {https://ui.adsabs.harvard.edu/abs/1990MNRAS.242..560L}
}

@ARTICLE{Dovciak2004,
       author = {{Dov{\v{c}}iak}, M. and {Karas}, V. and {Matt}, G.},
        title = "{Polarization signatures of strong gravity in active galactic nuclei accretion discs}",
      journal = {\mnras},
         year = 2004,
        month = dec,
       volume = {355},
       number = {3},
        pages = {1005-1009},
          doi = {10.1111/j.1365-2966.2004.08396.x},
archivePrefix = {arXiv},
       eprint = {astro-ph/0409356},
 primaryClass = {astro-ph},
}

@ARTICLE{Madau_LF2026,
       author = {{Madau}, Piero and {Maiolino}, Roberto},
        title = "{Little red dots as obscured little blue dots: Relative abundances, luminosities, and black hole masses}",
      journal = {\aap},
         year = 2026,
        month = nov,
       volume = {715},
          eid = {A15},
        pages = {A15},
          doi = {10.1051/0004-6361/202660815},
archivePrefix = {arXiv},
       eprint = {2605.05074},
 primaryClass = {astro-ph.GA},
       adsurl = {https://ui.adsabs.harvard.edu/abs/2026A&A...715A..15M}
}

@ARTICLE{MadauMaiolino2026,
       author = {{Madau}, Piero and {Maiolino}, Roberto},
        title = "{Little Red Dots as Obscured Little Blue Dots: A Super-Eddington Unification Model}",
      journal = {\aap},
         year = 2026,
        month = oct,
       volume = {714},
          eid = {A276},
        pages = {A276},
          doi = {10.1051/0004-6361/202659606},
archivePrefix = {arXiv},
       eprint = {2602.22386},
 primaryClass = {astro-ph.GA},
       adsurl = {https://ui.adsabs.harvard.edu/abs/2026A&A...714A.276M}
}

@ARTICLE{MadauWings2026,
       author = {{Madau}, Piero and {Maiolino}, Roberto and {Scholtz}, Jan and {D'Eugenio}, Francesco},
        title = "{Wings of little dots: Exponential broad lines from a stratified broad-line region}",
      journal = {\aap},
         year = 2026,
        month = oct,
       volume = {714},
          eid = {A186},
        pages = {A186},
          doi = {10.1051/0004-6361/202660251},
archivePrefix = {arXiv},
       eprint = {2604.04216},
 primaryClass = {astro-ph.GA},
       adsurl = {https://ui.adsabs.harvard.edu/abs/2026A&A...714A.186M}
}

@ARTICLE{Juod2026,
       author = {{Juod{\v{z}}balis}, Ignas and {Maiolino}, Roberto and {Baker}, William M. and {Lake}, Emma Curtis and {Scholtz}, Jan and {D'Eugenio}, Francesco and {Trefoloni}, Bartolomeo and {Isobe}, Yuki and {Tacchella}, Sandro and {Bunker}, Andrew J. and {Carniani}, Stefano and {Charlot}, St{\'e}phane and {Jones}, Gareth C. and {Parlanti}, Eleonora and {Perna}, Michele and {Rinaldi}, Pierluigi and {Robertson}, Brant and {{\"U}bler}, Hannah and {Venturi}, Giacomo and {Willott}, Chris},
        title = "{JADES: comprehensive census of broad-line AGN from reionization to cosmic noon revealed by JWST}",
      journal = {\mnras},
         year = 2026,
        month = mar,
       volume = {546},
       number = {3},
          eid = {stag086},
        pages = {stag086},
          doi = {10.1093/mnras/stag086},
archivePrefix = {arXiv},
       eprint = {2504.03551},
 primaryClass = {astro-ph.GA},
       adsurl = {https://ui.adsabs.harvard.edu/abs/2026MNRAS.546ag086J}
}

@ARTICLE{Brazzini2026,
       author = {{Brazzini}, M. and {D'Eugenio}, F. and {Maiolino}, R. and {Lyu}, J. and {DeCoursey}, C. and {{\"U}bler}, H. and {Ji}, X. and {Juod{\v{z}}balis}, I. and {Scholtz}, J. and {Jones}, G.~C. and {Hainline}, K. and {Dalla Bont{\`a}}, E. and {{\'e}rez-Gonz{\'a}lez}, P.~G. P and {Geris}, S. and {Harshan}, A. and {Feruglio}, C. and {Bischetti}, M. and {Mazzolari}, G. and {Rieke}, G. and {Alberts}, S. and {Trefoloni}, B. and {Carniani}, S. and {Parlanti}, E. and {Marconi}, A. and {Risaliti}, G. and {Ramos Almeida}, C. and {Rinaldi}, P. and {Perna}, M. and {Zamora}, S. and {Lamperti}, I. and {Venturi}, G. and {Cresci}, G. and {Bunker}, Andrew J. and {Ivey}, L.~R.},
        title = "{The Little Blue and Red Dots Rosetta Stones: Non-Gaussian broad lines, hot dust, and X-ray weakness}",
      journal = {\aap{}},
         year = 2026,
        month = jan,
         note = {in press},
          doi = {10.48550/arXiv.2601.22214},
archivePrefix = {arXiv},
       eprint = {2601.22214},
 primaryClass = {astro-ph.GA},
       adsurl = {https://ui.adsabs.harvard.edu/abs/2026arXiv260122214B}
}

@ARTICLE{Lin2026,
       author = {{Lin}, Xiaojing and {Fan}, Xiaohui and {Cai}, Zheng and {Bian}, Fuyan and {Liu}, Hanpu and {Sun}, Fengwu and {Ma}, Yilun and {Greene}, Jenny E. and {Strauss}, Michael A. and {Green}, Richard and {Lyu}, Jianwei and {Champagne}, Jaclyn B. and {Goulding}, Andy D. and {Inayoshi}, Kohei and {Jin}, Xiangyu and {Leung}, Gene C.~K. and {Li}, Mingyu and {Liu}, Weizhe and {Liu}, Yichen and {Mao}, Junjie and {Pudoka}, Maria Anne and {Tee}, Wei Leong and {Wang}, Ben and {Wang}, Feige and {Wu}, Yunjing and {Yang}, Jinyi and {Zhang}, Haowen and {Zhu}, Yongda},
        title = "{The Discovery of Little Red Dots in the Local Universe: Signatures of Cool Gas Envelopes}",
      journal = {\apj},
         year = 2026,
        month = feb,
       volume = {997},
       number = {2},
          eid = {364},
        pages = {364},
          doi = {10.3847/1538-4357/ae2bdf},
archivePrefix = {arXiv},
       eprint = {2507.10659},
 primaryClass = {astro-ph.GA},
       adsurl = {https://ui.adsabs.harvard.edu/abs/2026ApJ...997..364L}
}

@ARTICLE{Hainline2025,
       author = {{Hainline}, Kevin N. and {Maiolino}, Roberto and {Juod{\v{z}}balis}, Ignas and {Scholtz}, Jan and {{\"U}bler}, Hannah and {D'Eugenio}, Francesco and {Helton}, Jakob M. and {Sun}, Yang and {Sun}, Fengwu and {Robertson}, Brant and {Tacchella}, Sandro and {Bunker}, Andrew J. and {Carniani}, Stefano and {Charlot}, Stephane and {Curtis-Lake}, Emma and {Egami}, Eiichi and {Johnson}, Benjamin D. and {Lin}, Xiaojing and {Lyu}, Jianwei and {P{\'e}rez-Gonz{\'a}lez}, Pablo G. and {Rinaldi}, Pierluigi and {Silcock}, Maddie S. and {Venturi}, Giacomo and {Williams}, Christina C. and {Willmer}, Christopher N.~A. and {Willott}, Chris and {Zhang}, Junyu and {Zhu}, Yongda},
        title = "{An Investigation into the Selection and Colors of Little Red Dots and Active Galactic Nuclei}",
      journal = {\apj},
         year = 2025,
        month = feb,
       volume = {979},
       number = {2},
          eid = {138},
        pages = {138},
          doi = {10.3847/1538-4357/ad9920},
archivePrefix = {arXiv},
       eprint = {2410.00100},
 primaryClass = {astro-ph.GA},
       adsurl = {https://ui.adsabs.harvard.edu/abs/2025ApJ...979..138H}
}

@ARTICLE{Harikane2023AGN,
       author = {{Harikane}, Yuichi and {Zhang}, Yechi and {Nakajima}, Kimihiko and {Ouchi}, Masami and {Isobe}, Yuki and {Ono}, Yoshiaki and {Hatano}, Shun and {Xu}, Yi and {Umeda}, Hiroya},
        title = "{A JWST/NIRSpec First Census of Broad-line AGNs at z = 4-7: Detection of 10 Faint AGNs with M $_{BH}$ {}10$^{6}$-{}10$^{8}$ M $_{{\ensuremath{\odot}}}$ and Their Host Galaxy Properties}",
      journal = {\apj},
         year = 2023,
        month = dec,
       volume = {959},
       number = {1},
          eid = {39},
        pages = {39},
          doi = {10.3847/1538-4357/ad029e},
archivePrefix = {arXiv},
       eprint = {2303.11946},
 primaryClass = {astro-ph.GA},
       adsurl = {https://ui.adsabs.harvard.edu/abs/2023ApJ...959...39H}
}

@ARTICLE{Madau2026,
       author = {{Madau}, Piero},
        title = "{Chasing the light: Shadowing, collimation, and the super-Eddington growth of infant black holes in JWST broad-line AGNs}",
      journal = {\aap},
         year = 2026,
        month = apr,
       volume = {708},
          eid = {A116},
        pages = {A116},
          doi = {10.1051/0004-6361/202659244},
       adsurl = {https://ui.adsabs.harvard.edu/abs/2026A&A...708A.116M}
}

@ARTICLE{MaiolinoAGN,
       author = {{Maiolino}, Roberto and {Scholtz}, Jan and {Curtis-Lake}, Emma and {Carniani}, Stefano and {Baker}, William and {de Graaff}, Anna and {Tacchella}, Sandro and {{\"U}bler}, Hannah and {D'Eugenio}, Francesco and {Witstok}, Joris and {Curti}, Mirko and {Arribas}, Santiago and {Bunker}, Andrew J. and {Charlot}, St{\'e}phane and {Chevallard}, Jacopo and {Eisenstein}, Daniel J. and {Egami}, Eiichi and {Ji}, Zhiyuan and {Jones}, Gareth C. and {Lyu}, Jianwei and {Rawle}, Tim and {Robertson}, Brant and {Rujopakarn}, Wiphu and {Perna}, Michele and {Sun}, Fengwu and {Venturi}, Giacomo and {Williams}, Christina C. and {Willott}, Chris},
        title = "{JADES: The diverse population of infant black holes at 4 < z < 11: Merging, tiny, poor, but mighty}",
      journal = {\aap},
         year = 2024,
        month = nov,
       volume = {691},
          eid = {A145},
        pages = {A145},
          doi = {10.1051/0004-6361/202347640},
       adsurl = {https://ui.adsabs.harvard.edu/abs/2024A&A...691A.145M}
}

@ARTICLE{GoosmannGaskell2007,
       author = {{Goosmann}, Ren{\'e} W. and {Gaskell}, C. Martin},
        title = "{Modeling optical and UV polarization of AGNs. I. Imprints of individual scattering regions}",
      journal = {\aap},
         year = 2007,
        month = mar,
       volume = {465},
        pages = {129-145},
          doi = {10.1051/0004-6361:20053555}
}

@ARTICLE{Marin2014,
       author = {{Marin}, F.},
        title = "{A compendium of AGN inclinations with corresponding UV/optical continuum polarization measurements}",
      journal = {\mnras},
         year = 2014,
        month = jun,
       volume = {441},
       number = {1},
        pages = {551-564},
          doi = {10.1093/mnras/stu593},
archivePrefix = {arXiv},
       eprint = {1404.2417},
 primaryClass = {astro-ph.GA},
       adsurl = {https://ui.adsabs.harvard.edu/abs/2014MNRAS.441..551M}
}

@ARTICLE{Taylor2025_BHMF,
       author = {{Taylor}, Anthony J. and {Finkelstein}, Steven L. and {Kocevski}, Dale D. and {Jeon}, Junehyoung and {Bromm}, Volker and {Amor{\'\i}n}, Ricardo O. and {Arrabal Haro}, Pablo and {Backhaus}, Bren E. and {Bagley}, Micaela B. and {Banados}, Eduardo and {Bhatawdekar}, Rachana and {Brooks}, Madisyn and {Calabr{\`o}}, Antonello and {Ch{\'a}vez Ortiz}, {\'O}scar A. and {Cheng}, Yingjie and {Cleri}, Nikko J. and {Cole}, Justin W. and {Davis}, Kelcey and {Dickinson}, Mark and {Donnan}, Callum and {Dunlop}, James S. and {Ellis}, Richard S. and {Fern{\'a}ndez}, Vital and {Fontana}, Adriano and {Fujimoto}, Seiji and {Giavalisco}, Mauro and {Grazian}, Andrea and {Guo}, Jingsong and {Hathi}, Nimish P. and {Holwerda}, Benne W. and {Hirschmann}, Michaela and {Inayoshi}, Kohei and {Kartaltepe}, Jeyhan S. and {Khusanova}, Yana and {Koekemoer}, Anton M. and {Kokorev}, Vasily and {Larson}, Rebecca L. and {Leung}, Gene C.~K. and {Lucas}, Ray A. and {McLeod}, Derek J. and {Napolitano}, Lorenzo and {Onoue}, Masafusa and {Pacucci}, Fabio and {Papovich}, Casey and {P{\'e}rez-Gonz{\'a}lez}, Pablo G. and {Pirzkal}, Nor and {Somerville}, Rachel S. and {Trump}, Jonathan R. and {Wilkins}, Stephen M. and {Yung}, L.~Y. Aaron and {Zhang}, Haowen},
        title = "{Broad-line AGNs at 3.5 < z < 6: The Black Hole Mass Function and a Connection with Little Red Dots}",
      journal = {\apj},
         year = 2025,
        month = jun,
       volume = {986},
       number = {2},
          eid = {165},
        pages = {165},
          doi = {10.3847/1538-4357/add15b},
archivePrefix = {arXiv},
       eprint = {2409.06772},
 primaryClass = {astro-ph.GA},
       adsurl = {https://ui.adsabs.harvard.edu/abs/2025ApJ...986..165T}
}

@ARTICLE{Billand2026,
       author = {{Billand}, Jean-Baptiste and {Elbaz}, David and {Franco}, Maximilien and {Gentile}, Fabrizio and {Daddi}, Emanuele and {Giavalisco}, Mauro and {Kocevski}, Dale D. and {Lewis}, Joseph S. W. and {Magnelli}, Benjamin and {Sangalli}, Valentina and {Tarrasse}, Maxime},
        title = "{Do little red dots really form a distinct class of astronomical objects?}",
      journal = {arXiv e-prints},
         year = 2026,
        month = apr,
          eid = {arXiv:2604.11677},
        pages = {arXiv:2604.11677},
          doi = {10.48550/arXiv.2604.11677},
archivePrefix = {arXiv},
       eprint = {2604.11677},
 primaryClass = {astro-ph.GA},
       adsurl = {https://ui.adsabs.harvard.edu/abs/2026arXiv260411677B}
}

@INPROCEEDINGS{CohenMartel2001,
       author = {{Cohen}, Marshall H. and {Martel}, Andr\'{e} R.},
        title = "{Spectropolarimetry of Broad H$\alpha$ Lines and Geometry of the BLR}",
    booktitle = {Mass Outflow in Active Galactic Nuclei: New Perspectives},
       editor = {{Crenshaw}, D. M. and {Kraemer}, S. B. and {George}, I. M.},
       series = {Astronomical Society of the Pacific Conference Series},
       volume = {255},
        pages = {255},
         year = {2001},
archivePrefix = {arXiv},
       eprint = {astro-ph/0106168}
}

@ARTICLE{Moffat1969,
       author = {{Moffat}, A.~F.~J.},
        title = "{A Theoretical Investigation of Focal Stellar Images in the Photographic Emulsion and Application to Photographic Photometry}",
      journal = {Astronomy and Astrophysics},
         year = {1969},
       volume = {3},
        pages = {455--461}
}

@ARTICLE{Andersson2015,
       author = {{Andersson}, B.-G. and {Lazarian}, A. and {Vaillancourt}, J.~E.},
        title = "{Interstellar Dust Grain Alignment}",
      journal = {Annual Review of Astronomy and Astrophysics},
         year = {2015},
       volume = {53},
        pages = {501--539},
          doi = {10.1146/annurev-astro-082214-122414}
}

@ARTICLE{Marin2012,
       author = {{Marin}, F. and {Goosmann}, R.~W. and {Gaskell}, C.~M. and {Porquet}, D. and {Dov\v{c}iak}, M.},
        title = "{Modeling optical and UV polarization of AGNs. II. Polarization imaging and complex reprocessing}",
      journal = {Astronomy and Astrophysics},
         year = {2012},
       volume = {548},
          eid = {A121},
        pages = {A121},
          doi = {10.1051/0004-6361/201219751}
}

@ARTICLE{Marin2015,
       author = {{Marin}, F. and {Goosmann}, R.~W. and {Gaskell}, C.~M.},
        title = "{Modeling optical and UV polarization of AGNs. III. From uniform-density to clumpy regions}",
      journal = {Astronomy and Astrophysics},
         year = {2015},
       volume = {577},
          eid = {A66},
        pages = {A66},
          doi = {10.1051/0004-6361/201525628}
}

@ARTICLE{RojasLobos2018,
       author = {{Rojas Lobos}, P.~A. and {Goosmann}, R.~W. and {Marin}, F. and {Savi\'c}, D.},
        title = "{Modeling optical and UV polarization of AGNs. IV. Polarization timing}",
      journal = {Astronomy and Astrophysics},
         year = {2018},
       volume = {611},
          eid = {A39},
        pages = {A39},
          doi = {10.1051/0004-6361/201731331}
}

@ARTICLE{Marin2018,
       author = {{Marin}, F. and {Dov\v{c}iak}, M. and {Muleri}, F. and {Kislat}, F.~F. and {Krawczynski}, H.~S.},
        title = "{Predicting the X-ray polarization of type 2 Seyfert galaxies}",
      journal = {Monthly Notices of the Royal Astronomical Society},
         year = {2018},
       volume = {473},
        pages = {1286--1316},
          doi = {10.1093/mnras/stx2382}
}

@ARTICLE{GoosmannMatt2011,
       author = {{Goosmann}, R.~W. and {Matt}, G.},
        title = "{Spotting the misaligned outflows in NGC 1068 using X-ray polarization}",
      journal = {Monthly Notices of the Royal Astronomical Society},
         year = {2011},
       volume = {415},
        pages = {3119--3128},
          doi = {10.1111/j.1365-2966.2011.18923.x}
}

\begin{appendix}

\section{Model geometries}
\label{app:model_geometry}

Figure~\ref{fig:model_geometry} summarizes the external-scattering
alternatives considered below; the baseline disk-polarization geometry is
shown in Figure~\ref{fig:baseline_geometry}.
\setcounter{figure}{0}
\begin{figure}[!ht]
\centering
\includegraphics[width=\columnwidth]{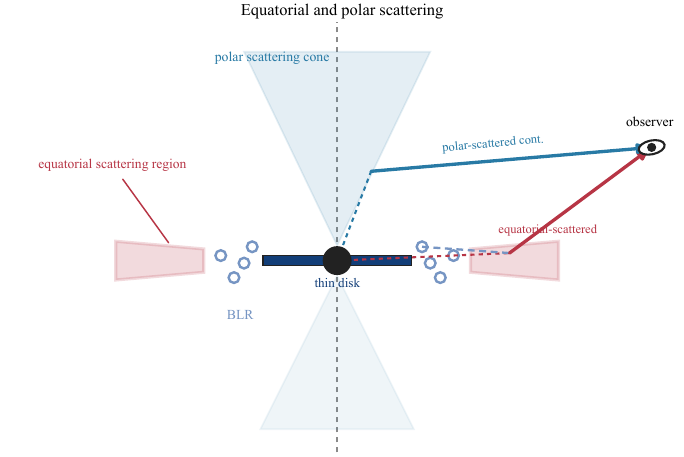}
\caption{Schematic external-scattering geometry. A narrow flared equatorial
electron-scattering annulus with a $5^\circ$ half-opening angle surrounds the BLR,
while a polar scattering cone redirects radiation into the line of sight.
The equatorial calculations below adopt the geometrically thin-ring limit of
this configuration. Dashed segments indicate incident paths, and
solid colored segments indicate the corresponding scattered or observed
paths. The drawing is schematic and not to scale.}
\label{fig:model_geometry}
\end{figure}
\section{A conical toy model}
\label{app:conical_toy}

At each surface element $(r,\phi)$, the polarization is perpendicular to the
local meridian plane defined by the surface normal $\hat{\boldsymbol{n}}$ and
the observer direction $\hat{\boldsymbol{k}}$. As $\phi$ runs from $0$ to
$2\pi$, the polarization angle $\chi(r,\phi;i)$ rotates on the sky and
$\cos 2\chi$ changes sign. Face-on, $\mu$ and $p(\mu)$ are independent of
$\phi$ and the tangential polarization cancels exactly in the azimuthal
integral. Inclination makes the projected weights and directions
azimuth-dependent, leaving a finite residual rather than the local
Chandrasekhar polarization of any one surface element.

This behavior can be illustrated by considering a circular ring on a
conical funnel wall of half-opening angle $\Theta$, measured from the
symmetry axis. We neglect non-local self-shadowing. For the surface normal
directed into the funnel, we write
\begin{equation}
\hat{\boldsymbol{n}}
=
(-\cos\Theta\cos\phi,\,
 -\cos\Theta\sin\phi,\,
 \sin\Theta),
\end{equation}
while the observer direction is
\begin{equation}
\hat{\boldsymbol{k}}
=
(\sin i,\,0,\,\cos i).
\end{equation}
The local emission-angle cosine is therefore
\begin{equation}
\mu(\phi;i,\Theta)
=
\hat{\boldsymbol{n}}\cdot\hat{\boldsymbol{k}}
=
\sin\Theta\cos i-\cos\Theta\sin i\cos\phi .
\label{eq:mu_cone}
\end{equation}
For $i>\Theta$, $\mu$ becomes negative near $\phi=0$. These elements are
back-facing and are excluded by the local visibility condition $\mu>0$.
This is distinct from non-local self-shadowing, in which the ray from an
otherwise visible element to the observer intersects another part of the
funnel; the latter effect is omitted in this toy model.

At exactly face-on inclination, $\mu=\sin\Theta$ is independent of azimuth,
so every surface element has the same local Chandrasekhar polarization
$p(\sin\Theta)$. The projected polarization direction nevertheless rotates
around the ring. For $i>0$, choosing the projected symmetry axis as the
reference direction on the sky gives
\begin{equation}
\cos 2\chi
=
2\,\frac{\cos^2\Theta\,\sin^2\phi}{1-\mu^2}-1 .
\label{eq:cos2chi_cone}
\end{equation}
Taking the face-on limit, $i\rightarrow0$, yields
\begin{equation}
\cos 2\chi
=
2\sin^2\phi-1
=
-\cos 2\phi ,
\end{equation}
and hence
\begin{equation}
\begin{aligned}
Q_{\nu,{\rm sky}}
&\propto
p(\sin\Theta)I_\nu(r,\sin\Theta)\sin\Theta
\int_0^{2\pi}\cos 2\chi\,{\rm d}\phi
=0,\\
U_{\nu,{\rm sky}}
&=0 .
\end{aligned}
\end{equation}
Thus the local polarization may be several percent even though the
surface-integrated polarization vanishes exactly by azimuthal cancellation.

For a small but finite inclination satisfying $i<\Theta$, all azimuths
remain locally visible and
\begin{equation}
\begin{aligned}
\mu(\phi;i,\Theta)
&\simeq \sin\Theta-\cos\Theta\,i\cos\phi\\
&\quad-\frac{1}{2}\sin\Theta\,i^2+\mathcal O(i^3),
\end{aligned}
\qquad i\ll1 .
\end{equation}
Both the projected weights and the sky-projected polarization directions
then acquire an azimuthal dependence. To isolate this geometric effect, we
hold $p(\mu)$ and ${\cal H}(\mu)$ fixed at their face-on values. Reflection symmetry
gives $U_{\nu,{\rm sky}}=0$. Defining the observed polarization fraction of
the conical ring as
\begin{equation}
p_{\rm obs}
\equiv
\frac{
\sqrt{Q_{\nu,{\rm sky}}^2+U_{\nu,{\rm sky}}^2}
}{
f_{\nu,{\rm obs}}
},
\end{equation}
we obtain
\begin{equation}
p_{\rm obs}
=
p(\sin\Theta)
\left|
\frac{
\displaystyle\int_0^{2\pi}
\mu(\phi;i,\Theta)\cos 2\chi\,{\rm d}\phi
}{
\displaystyle\int_0^{2\pi}
\mu(\phi;i,\Theta)\,{\rm d}\phi
}
\right|.
\end{equation}
Expanding to the first non-vanishing order in inclination gives
\begin{equation}
p_{\rm obs}
\simeq
\frac{3}{4}\,p(\sin\Theta)\,i^2,
\qquad i\ll1,
\label{eq:pobs_smalli}
\end{equation}
with $i$ expressed in radians. The net polarization seen by the observer is
therefore quadratically suppressed at low inclination even though every
surface element retains a finite local Chandrasekhar polarization. In this
simplified limit, the funnel opening angle enters only through the local
value $p(\sin\Theta)$. At larger inclinations, this cancellation becomes progressively
less effective until local visibility and, in the full disk calculation,
self-shadowing become important.

\section{Full-Stokes Monte Carlo test of the funnel approximation}
\label{app:stokes_benchmark}

The calculation in Section~2 uses scalar Lambertian wall coupling and applies
a local Chandrasekhar law at the final emitting surface, without preserving
the direction or polarization state between encounters. We test this closure
with the public STOKES v1.2 Monte Carlo code
\citep{GoosmannGaskell2007,Marin2012,Marin2015,RojasLobos2018,Marin2018},
which applies the Thomson Mueller matrix at each interaction and follows
$(I,Q,U,V)$ through arbitrarily many scatterings. We added a conical-shell
geometry with analytic ray--boundary intersections and a wall-distributed source without changing
the scattering physics or Stokes updates. The implementation passes
analytic-intersection and deterministic-reference tests.

\begin{figure}[!htb]
\centering
\includegraphics[width=\columnwidth]{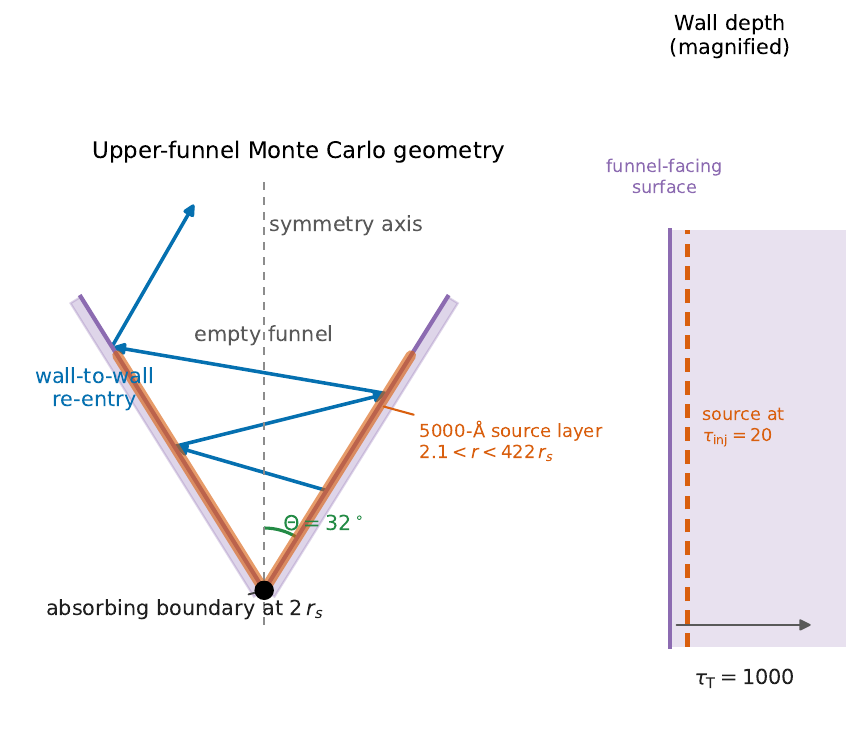}
\caption{Full-STOKES benchmark geometry. By symmetry, we simulate only the
upper funnel, with half-opening angle $\Theta=32^\circ$, a pure-electron
conical wall of normal Thomson depth $\tau_{\rm T}=1000$, and a central
absorbing boundary at $2\,r_s$. The cone
extends to a spherical radius of $1000\,r_s$ ($r=530\,r_s$, $h=848\,r_s$),
providing a simplified representation of the realistic thick-torus funnel.
Initially unpolarized photons are injected isotropically at
$\tau_{\rm inj}=20$ and sampled over the wall with weight
$\mathcal F_\nu^{\rm local}(r)\,r\,{\rm d}r\,{\rm d}\phi$. They may propagate
between and scatter repeatedly from the opposing funnel walls while the full Stokes vector is
updated at every Thomson scattering. The source-layer offset is exaggerated.}
\label{fig:stokes_geometry}
\end{figure}

Figure~\ref{fig:stokes_geometry} shows the setup. By symmetry, we simulate only
the upper funnel. Here $r_s\equiv2GM_{\rm BH}/c^2$ is the Schwarzschild radius.
The wall is a straight cone of half-opening angle $\Theta=32^\circ$, extending
from the apex to a spherical radius of $1000\,r_s$. It is a constant-density,
pure-electron-scattering atmosphere with total normal Thomson depth
$\tau_{\rm T}=1000$; at fixed $\tau_{\rm T}$, transfer is controlled by optical
depth rather than geometric thickness. A central spherical boundary of radius
$2\,r_s$ absorbs photons that intersect it. Photons may cross the empty funnel,
enter the opposite wall, and scatter again.

Initially unpolarized photons are injected isotropically at normal depth
$\tau_{\rm inj}=20$ below the funnel-facing surface. Their probability per
surface element is proportional to the exact local, pre-irradiation
$5000\,$\AA\ flux $\mathcal F_\nu^{\rm local}(r)$ of the best-fitting
$M_{\rm BH}=10^{7.5}\,M_\odot$, $\dot m=31.7$ solution. They are sampled with
weight $\mathcal F_\nu^{\rm local}(r)\,r\,{\rm d}r\,{\rm d}\phi$, where $r$ is
cylindrical radius; the constant conical slope contributes only an overall
area factor. Injection spans $2.1<r/r_s<422$. The benchmark retains the
numerical model's local emissivity but replaces its curved photosphere, whose
maximum height is $\simeq325\,r_s$, with a controlled cone. It tests the
transfer closure, not the disk geometry.

The reference calculation uses the same cone, source profile, and detector
bins, sums scalar Lambertian irradiation to all orders, and applies the
paper's Chandrasekhar $H$-function and local polarization law. We compare the
emergent intensity and signed $Q/I$ for bins wholly inside the funnel,
$i<\Theta$; positive $Q$ is parallel to the projected axis. Axisymmetry
requires $U/I=0$ at every inclination and $Q/I=U/I=0$ at $i=0$, so $U$ is only
a symmetry/noise diagnostic. The calculation comprises 400 statistically
independent Monte Carlo realizations, each with 25,000 photons and a different
random seed ($10^7$ photons total). We sum $(I,Q,U,V)$
within each of four equal-solid-angle intervals before forming normalized
ratios; uncertainties are the standard errors determined from the scatter
among the 400 realizations.

At any fixed inclination, the angle between the local wall normal and the
observer varies with azimuth around the cone. Equation~(\ref{eq:mu_cone})
shows that the corresponding emission-angle cosine $\mu$ spans
$0.40$--$0.65$, $0.30$--$0.73$, $0.19$--$0.80$, and $0.08$--$0.86$ for the
four plotted inclination bins, respectively. The local Chandrasekhar closure would therefore
assign polarization fractions to individual surface elements
$p(\mu)=1.2$--$2.9\%$, $0.9$--$4.0\%$, $0.6$--$5.6\%$, and $0.4$--$8.3\%$,
respectively. Their projected-intensity-weighted unsigned means are $1.9\%$,
$1.8\%$, $1.6\%$, and $1.4\%$. The much smaller integrated polarization is an
azimuthal-cancellation effect: $\cos 2\chi$ changes sign around the funnel, so
oppositely signed contributions to $Q$ largely cancel, while reflection
symmetry cancels $U$.

Figure~\ref{fig:stokes_q_comparison} compares the resulting signed polarization
with the paper's approximation. At the centers of the four detector bins wholly
inside the funnel, $i=8.5^\circ$, $14.8^\circ$, $21.0^\circ$, and
$27.3^\circ$, STOKES gives $Q/I=0.04\pm0.09\%$, $0.05\pm0.09\%$,
$0.22\pm0.06\%$, and $0.31\pm0.07\%$, respectively. The corresponding
approximation gives $0.03\%$, $0.09\%$, $0.20\%$, and $0.34\%$. The
differences are $0.2$, $0.5$, $0.4$, and $0.5\sigma$; the agreement is
excellent. The first two bins are nondetections consistent with both zero and
the approximation, whereas the two higher-inclination bins are positive
detections. Both calculations show that the polar signal is strongly suppressed
by azimuthal cancellation. Multiple scatterings within the optically thick
funnel walls drive the radiation field toward the diffusion regime. In this
limit, the radiation largely loses memory of its incident direction and
polarization, and the escaping polarization is determined primarily by the
final scatterings near the photosphere. This explains why the local
Chandrasekhar approximation accurately reproduces the full STOKES
calculation. The agreement extends to intensity: after
normalization over these bins, the STOKES-to-approximation ratios are $1.03$,
$1.02$, $1.00$, and $0.98$. The symmetry diagnostic is
$U/I=-0.07\pm0.10\%$, $-0.01\pm0.09\%$,
$+0.02\pm0.07\%$, and $-0.12\pm0.07\%$; every value is within $1.9\sigma$
of zero. Within the Monte Carlo precision of this benchmark, the scalar wall coupling plus
local Chandrasekhar escape law reproduces both the angular intensity and the
small funnel polarization obtained by the full-Stokes multiple-scattering
calculation.

The straight benchmark cone has a half-opening angle $\Theta=32^\circ$, so
only sightlines with $i<\Theta$, for which the inner funnel is directly
visible, provide a controlled test of the closure. Extending this artificial
cone to $i\simeq90^\circ$ would not test the relevant high-inclination signal
of the numerical disk: there the inner funnel is self-obscured and the
observed polarization is dominated by the curved outer rim, which is absent
from the benchmark. A full transfer calculation for that geometry would
require implementing the numerical photosphere itself.

\begin{figure}[!htb]
\centering
\includegraphics[width=\columnwidth]{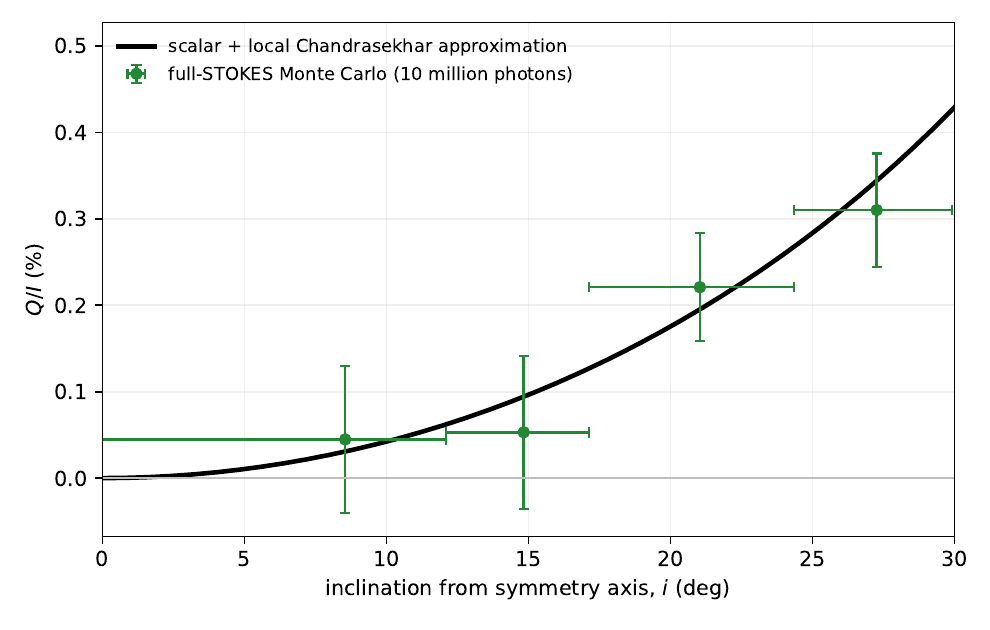}
\caption{$Q/I$ from the $10^7$-photon full-STOKES calculation and the paper's
scalar wall-coupling plus local Chandrasekhar approximation. The black curve is
continuous; only the STOKES results are detector-binned. Positive $Q$ is
parallel to the projected axis. Vertical bars are $1\sigma$ uncertainties from
400 statistically independent 25,000-photon realizations; horizontal bars show the
equal-solid-angle bin widths.}
\label{fig:stokes_q_comparison}
\end{figure}

\section{An equatorial-scattering model}
\label{app:equatorial_scattering}

We test whether a conventional equatorial region can instead produce the
continuum and broad-line polarization. In the standard type~1 Seyfert
picture, an unpolarized central continuum and rotating, disk-like BLR
illuminate a coplanar electron-scattering region between the BLR and the
inner edge of the dusty torus
\citep{Smith2002,Smith2005,Lira2020}. The continuum polarization
is then parallel to the projected rotation axis, while the approaching and
receding sides of the BLR produce opposite rotations of the polarization
position angle across a broad line and enhanced polarization in its wings.
These signatures have also been explored with Monte Carlo polarized
radiative-transfer calculations \citep{Lira2020}.

Our deliberately simple realization consists of a thin H$\alpha$-emitting
ring of cylindrical radius $r_{\rm BLR}$, surrounded by a stationary,
optically thin electron-scattering ring at cylindrical radius $r_{\rm sc}$.
Both rings are complete and
azimuthally symmetric; no obscuration or preferred scattering sector is
introduced. The direct continuum and broad H$\alpha$ are both taken to be
intrinsically unpolarized. For a ray scattered through angle $\xi$, the
Thomson polarization fraction is
$(1-\cos^2\xi)/(1+\cos^2\xi)$.
Let $\phi$ be the azimuth of a scattering element measured from the projected
minor axis. Its continuum scattering angle satisfies
\begin{equation}
\cos\xi=\sin i\cos\phi.
\label{eq:eqscat_angle}
\end{equation}
Choosing positive Stokes $Q$ along the projected rotation axis, the elemental
continuum Stokes fluxes, apart from a common normalization, are
\begin{align}
{\rm d}I_{\rm sc}&\propto 1+\sin^2i\cos^2\phi,\nonumber\\
{\rm d}Q_{\rm sc}&\propto
\sin^2\phi-\cos^2i\cos^2\phi,\\
{\rm d}U_{\rm sc}&\propto-\cos i\sin2\phi.\nonumber
\label{eq:eqscat_stokes}
\end{align}
Azimuthal integration therefore gives $U_{\rm sc}=0$ but leaves a finite
polarization
\begin{equation}
p_{\rm sc}=\frac{\sin^2 i}{2+\sin^2 i}.
\label{eq:eqscat_psc}
\end{equation}
The cancellation is complete only for a face-on system. At $i>0$, elements on
the projected major axis scatter through $90^\circ$ and dominate the polarized
flux over their minor-axis counterparts. Following \citet{Smith2005}, each
scattering element is assigned the same broad-line profile and position-angle
rotation relative to its local continuum angle; these elemental profiles are
then rotated and Thomson-weighted around the complete ring. For a rotating
BLR, the resulting velocity-resolved Stokes profiles obey
\begin{equation}
Q_{\rm sc}(-v)=Q_{\rm sc}(v),
\qquad
U_{\rm sc}(-v)=-U_{\rm sc}(v),
\label{eq:eqscat_velocity_symmetry}
\end{equation}
and hence produce equal and opposite position-angle rotations in the blue and
red wings even though the ring is axisymmetric. A Gaussian local velocity
dispersion and the instrumental line-spread function are applied before
averaging over the adaptive wavelength bins. This construction implements the
narrow-ring benchmark of \citet{Smith2005}, rather than a full radiative-
transfer model.

The wavelength and velocity sampling is specific to J1025+1402. We use its
observed broad H$\alpha$ profile and include the direct narrow H$\alpha$
component, with intrinsic ${\rm FWHM}=92\,\mathrm{km\,s^{-1}}$
\citep{Ji2026}, as unpolarized intensity dilution. Appendices~D--F are
deliberately optically thin, deterministic single-scattering null tests of
standard AGN geometries. They do not attempt an exhaustive radiative-transfer
fit. A realistic STOKES calculation would require presently unconstrained
optical depths, density and radial profiles, vertical extent, clumpiness,
occultation, and electron--dust mixtures. At approximately $10^7$ photons per
configuration, exploring that parameter space would require a dedicated
high-performance-computing grid beyond the scope justified by the present
data.

The continuum constrains only the product of the scattered fraction and the
net polarization of the scattered component. Let $f_{\rm sc}$ be the fraction
of the observed continuum supplied by scattered light and $p_{\rm sc}$ its
polarization fraction, given by Equation~(\ref{eq:eqscat_psc}). Their product
gives
\begin{equation}
p_{\rm cont}=f_{\rm sc}p_{\rm sc}.
\label{eq:eqscat_degeneracy}
\end{equation}
Thus $f_{\rm sc}$ and $p_{\rm sc}$ are degenerate: the data determine their
product, whereas $p_{\rm sc}$ follows only after a scattering geometry is
specified. For each geometry we therefore fit the continuum $Q$ and $U$
fluxes alone, with free scattered amplitude $f_{\rm sc}p_{\rm sc}$ and angle
$\psi_{\rm axis}$, and then predict the H$\alpha$ Stokes
profiles with the same scattering normalization. No independent broad-line
polarization amplitude is introduced.

We scan $i=35^\circ$, $45^\circ$, $60^\circ$, and $70^\circ$,
$r_{\rm sc}/r_{\rm BLR}=1.5$, 2, and 5, and circular speeds from 500 to
$2000\,\mathrm{km\,s^{-1}}$. Every geometry gives the same continuum solution,
$p_{\rm cont}=1.5\%$ and $\psi_{\rm axis}=13^\circ$. The lowest total
$\chi^2$ occurs for $i=70^\circ$, $r_{\rm sc}/r_{\rm BLR}=5$,
$v_\phi=1000\,\mathrm{km\,s^{-1}}$. In that geometry,
Equation~(\ref{eq:eqscat_psc}) gives $p_{\rm sc}=31\%$, implying
$f_{\rm sc}=4.9\%$ through
Equation~(\ref{eq:eqscat_degeneracy}).
The model gives $\chi^2=1056$ for 133 nominal degrees of freedom, using the
same 138 Stokes-flux data values and full covariance matrices as the main
analysis, compared with $\chi^2=210.8$ for the
disk-plus-dichroic model. Its continuum contribution is acceptable
($\chi^2_{\rm cont}=78$), but H$\alpha$ alone contributes
$\chi^2_{\mathrm{H}\alpha}=978$. Moving the scatterer outward reduces the
position-angle rotation because the BLR becomes increasingly point-like as
seen by the scatterer. Nevertheless, the modeled H$\alpha$ window spans
approximately $25^\circ$, $17^\circ$, and $6^\circ$ in $\vartheta$ for
$r_{\rm sc}/r_{\rm BLR}=1.5$, 2, and 5, respectively, and all three cases
overpredict the line polarization. Thus, varying the scatterer-to-BLR radius
changes the amplitude of the predicted position-angle rotation but does not
resolve the disagreement with the observed weak H$\alpha$ Stokes structure.

This comparison does not exclude every external-scattering geometry. Partial
obscuration by the atmosphere of the dusty torus could produce an asymmetric
scattering arc \citep{CohenMartel2001,Smith2005}; a vertically or radially
extended scatterer, multiple scattering, or an additional polar component
could likewise alter both the line strength and its
position-angle profile \citep{Smith2005,Lira2020}. It does show that the
simplest complete-ring equatorial model, in which the same scatterer polarizes the
continuum and broad H$\alpha$, is inconsistent with the data. Allowing an
arbitrary independent line normalization would remove the defining
prediction and would not constitute the standard single-scatterer test.

\section{Equatorial scattering plus dichroic transmission}
\label{app:equatorial_dichroic}

We next test whether the equatorial ring can reproduce the data when combined
with the foreground dichroic screen used in the main analysis. The intrinsic
continuum and broad H$\alpha$ emission remain unpolarized, and the same
complete ring scatters both components. After the direct and scattered fluxes
are combined, they pass through a uniform dichroic screen. At each wavelength,
the equatorial-scattering and dichroic Stokes vectors are summed
deterministically in $Q$--$U$ space before the normalized polarization and
position angle are formed. To first order in
the small polarization fractions, the normalized Stokes parameters are
\begin{align}
q_{\rm mod}(\lambda)
&=q_{\rm eq}(\lambda)
 +p_{\rm dic}(\lambda)\cos2\psi_{\rm dic},\nonumber\\
u_{\rm mod}(\lambda)
&=u_{\rm eq}(\lambda)
 +p_{\rm dic}(\lambda)\sin2\psi_{\rm dic},
\label{eq:eqdic_stokes}
\end{align}
where $q_{\rm eq}$ and $u_{\rm eq}$ include dilution by the direct flux and
the velocity-dependent broad-line signal of Appendix~\ref{app:equatorial_scattering}.
The dichroic amplitude follows the Serkowski law in
Equation~(\ref{eq:J1025_Serkowski}), with the same fixed $\lambda_{\rm max}$
and $K$. We retain the defining constraint of the standard model: the
equatorial-scattering normalization is the same for the continuum and broad
H$\alpha$. No independent line-polarization amplitude or narrow-line
component is introduced.

We scan the same $(i,r_{\rm sc}/r_{\rm BLR},v_\phi)$ grid as in
Appendix~\ref{app:equatorial_scattering} and jointly fit the equatorial
continuum amplitude and position angle, $p_{\rm eq}$ and $\psi_{\rm eq}$,
together with $p_{\rm max}$ and $\psi_{\rm dic}$. The best grid point has
$i=70^\circ$, $r_{\rm sc}/r_{\rm BLR}=5$, and
$v_\phi=2000\,\mathrm{km\,s^{-1}}$. It gives
$p_{\rm eq}=0.38\%$ at $\psi_{\rm eq}=28^\circ$ and
$p_{\rm max}=0.74\%$ at $\psi_{\rm dic}=-3^\circ$. Their vector sum yields
a mean continuum polarization of $0.97\%$ at $7^\circ$. For this geometry,
$p_{\rm sc}=31\%$ and the inferred scattered-continuum fraction is
$f_{\rm sc}=1.2\%$.
The fit gives $\chi^2=450$ for 131 nominal degrees of freedom, with
$\chi^2_{\rm cont}=191$ and $\chi^2_{\mathrm{H}\alpha}=259$. This is a large
improvement over equatorial scattering alone ($\chi^2=1056$), because the
dichroic and scattering vectors partially cancel across H$\alpha$, but it
remains substantially worse than the disk-plus-dichroic model
($\chi^2=210.8$). Increasing $p_{\rm eq}$ would bring the continuum closer to
the observed $\simeq1.5\%$ polarization but would restore excessively strong
polarized wings; reducing it weakens the line structure but leaves the
continuum underpolarized. The best fit is therefore a compromise that cannot
match both observables simultaneously (Figure~\ref{fig:equatorial_dichroic}).

The discrepancy could be reduced by allowing the scattered H$\alpha$
normalization to be lower than that of the scattered continuum, or by placing
the scatterer interior to most of the BLR. Either change, however, abandons
the standard Smith geometry, in which a scattering layer outside the BLR
intercepts both the central continuum and broad-line radiation. The combined
test therefore disfavors equatorial ring scattering plus dichroic
transmission as a replacement for intrinsic disk polarization.
As emphasized in Appendix~\ref{app:equatorial_scattering}, this is a
single-scattering null test and not an exhaustive exclusion of more complex
equatorial transfer.

\section{Equatorial and polar scattering}
\label{app:equatorial_polar}

We finally test the standard Seyfert two-region picture, in which a compact
equatorial region inside the dusty torus and an external polar cone scatter
unpolarized continuum and BLR radiation \citep{Smith2002,Smith2004,Smith2005}.
The regions share a symmetry axis. For complete, axisymmetric scatterers, reflection
symmetry gives zero continuum Stokes $U$ in the axis-aligned frame, while the
equatorial and polar components have opposite signs of Stokes $Q$. Their
continuum polarization angles are therefore orthogonal in this idealized
geometry: the equatorial component lies at $\psi_{\rm axis}$ and the polar
component at $\psi_{\rm axis}+90^\circ$.
Individual elements of the polar cone span a range of scattering planes;
this orthogonality applies only after azimuthal integration and
would not hold exactly for a clumpy, tilted, or partially obscured cone
\citep{GoosmannMatt2011}.

The equatorial component is the complete ring of
Appendix~\ref{app:equatorial_scattering}. The polar component is represented
by an azimuthally complete annulus on the visible upper cone, at spherical
radius $R_{\rm pol}$ and half-opening angle $\Theta_{\rm pol}$. We calculate
its Stokes fluxes by deterministic single-scattering integration over the
polar annulus and the rotating H$\alpha$-emitting ring. If $f_{\rm eq}$ and
$f_{\rm pol}$ are the fractions of the observed line-free continuum supplied
by the two scattered components, their oppositely signed polarized
contributions subtract in the axis-aligned frame, where $U=0$, before rotation
by $\psi_{\rm axis}$ into the celestial frame. The remaining
fraction $1-f_{\rm eq}-f_{\rm pol}$ is direct unpolarized continuum. Both
scattering regions intercept the same continuum and broad-line spectrum, so
their H$\alpha$ normalizations are tied to their continuum normalizations.
No independent broad-line amplitude, dichroic component, or narrow-line
polarization is introduced; direct broad and narrow emission dilutes the
scattered Stokes flux through the observed total intensity.

We scan the inclinations and equatorial radii of
Appendix~\ref{app:equatorial_scattering}, together with
$R_{\rm pol}/r_{\rm BLR}=2$, 5, and 10,
$\Theta_{\rm pol}=20^\circ$, $30^\circ$, and $45^\circ$, and
$v_\phi=500$--$2000\,\mathrm{km\,s^{-1}}$. At each grid point we jointly fit
the total scattered-continuum fraction, the fraction supplied by the polar
region, and $\psi_{\rm axis}$. The best solution lies at
$i=70^\circ$, $R_{\rm pol}/r_{\rm BLR}=2$,
$\Theta_{\rm pol}=20^\circ$, and
$v_\phi=2000\,\mathrm{km\,s^{-1}}$. It drives the equatorial fraction to
zero, so the nominal two-region model collapses to the nested polar-only
case. The polar component supplies $1.7\%$ of the continuum; its intrinsic
single-scattering polarization is $63\%$, and the remaining $98.3\%$ is
direct unpolarized light. The resulting continuum polarization is only
$1.1\%$ at $7^\circ$, below the observed $\simeq1.5\%$ at
$13^\circ$.

This suppression of the equatorial component does not
conflict with the success of the dichroic screen in the baseline model. The
screen assigns the transmitted broad H$\alpha$ emission a single position
angle, $\psi_{\rm dic}$; the variation across the line arises mainly because
the disk-polarized continuum is progressively diluted as the line fraction
increases, so that the net angle moves toward $\psi_{\rm dic}$. Equatorial
scattering instead maps BLR Doppler velocity onto scattering geometry: the
blue and red sides sample different emitting locations and scattering planes,
producing velocity-dependent Stokes structure and a resolved position-angle
swing. The data require an overall continuum-to-broad-line angle offset but
do not show this kinematic structure, so the dichroic component improves the
fit while the complete equatorial ring is driven to zero.
This boundary solution of the full equatorial-plus-polar model gives
$\chi^2=461$ for 130 nominal degrees
of freedom, with $\chi^2_{\rm cont}=164$ and
$\chi^2_{\mathrm{H}\alpha}=297$ (Figure~\ref{fig:equatorial_polar}). It is
substantially worse than the disk-plus-dichroic model ($\chi^2=210.8$). The fit
reduces the polar-scattering amplitude to avoid overpolarizing broad
H$\alpha$, thereby underpredicting the continuum. Adding the equatorial ring
cannot repair this conflict because its velocity-dependent polarized wings
and position-angle rotation add line structure that is not present in the
data. Thus the standard tied equatorial-plus-polar geometry does not provide
a viable replacement for intrinsic disk polarization in J1025+1402.
The methodological limitations stated in
Appendix~\ref{app:equatorial_scattering} apply here as well: a clumpy,
optically thick, multiple-scattering polar region could produce different
Stokes structure, but testing such a model requires a much larger and poorly
constrained radiative-transfer grid.

\section{Radial origin of the signal and general-relativistic effects}
\label{app:gr_origin}

General-relativistic propagation can modify polarization through frequency
shifts, aberration, light bending, and parallel transport
\citep{Laor1990,Dovciak2004}, particularly when much of the Stokes flux
originates in the strong-field region. Rather than perform Kerr ray tracing,
we measure the fraction of optical/UV signal in the fitted Newtonian model
carried by photons born within a conservative boundary $r=20r_s$, where
$r_s\equiv2GM_{\rm BH}/c^2$. Since $r$ is cylindrical, the true distance
$(r^2+h^2)^{1/2}\geq r$, so this cut includes some emission produced beyond
$20r_s$.

The scalar wall-transfer equation is linear. Defining
$K_{kj}\equiv B_{kj}/\pi$, its solution may be decomposed by the local birth
radius of the photons,
\begin{align}
\boldsymbol{\mathcal F}_{\nu}
 &= (\boldsymbol{1}-\boldsymbol K)^{-1}
    \boldsymbol{\mathcal F}^{\rm local}_{\nu},\\
\boldsymbol{\mathcal F}^{<20}_{\nu}
 &= (\boldsymbol{1}-\boldsymbol K)^{-1}
    \left[H(20r_s-r)\,
    \boldsymbol{\mathcal F}^{\rm local}_{\nu}\right],
\label{eq:gr_origin_decomposition}
\end{align}
where $H$ is the Heaviside function. This construction counts a photon by its
birth radius even if it subsequently undergoes repeated wall-to-wall
reflection and finally escapes from a much larger radius. This birth-radius
tag remains relevant when assessing sensitivity to gravitational redshift:
within our conservative, frequency-preserving Thomson treatment, a wall
encounter redirects a photon but does not erase the frequency shift inherited
from its emission radius. The calculation does not follow curved photon
trajectories or their effects on wall illumination, self-shadowing, and
polarization-vector transport; treating those effects requires fully
relativistic polarized ray tracing.
We propagate the two source components through the same self-shadowing
calculation and apply the same Chandrasekhar $H$-function angular-intensity
and local polarization laws as in the main model. The unpolarized outer thin
disk is included in the denominator of the received intensity.

Figure~\ref{fig:gr_origin_fraction} shows, at the two wavelengths used in
Figure~\ref{fig:pol_decomp}, the fraction of received spectral intensity
carried by photons born at $r<20r_s$, including all subsequent scalar wall
reflections. Polarization requires a
different radial tag. Because each wall encounter in our closure discards the
incident Stokes state, the relevant strong-field location is the final
polarizing surface. To avoid the singular behavior produced when the net
signed $Q$ nearly cancels, we show the absolute-$Q$ fraction
\begin{equation*}
\frac{\sum_{k:r_k<20r_s} |\mathrm dQ_{\lambda,k}|}
     {\sum_k |\mathrm dQ_{\lambda,k}|}.
\end{equation*}
This includes a photon born elsewhere but last scattered inside $20r_s$, and
excludes one born inside $20r_s$ but subsequently reprocessed at a larger
radius. A full polarized-transfer calculation would preserve more memory of
the incident Stokes vector than this local Chandrasekhar closure.

At $i=65^\circ$, inner-born photons contribute $1.1\%$ and $0.63\%$ of the
received intensity at $1500\,$\AA\ and $6500\,$\AA; at $70^\circ$, these
fractions fall to $0.66\%$ and $0.39\%$. At both inclinations and wavelengths,
the fraction of $|\mathrm dQ|$ last polarized inside $20r_s$ is zero: the
self-shadowing tangents at $r\simeq210r_s$ and $240r_s$ hide the strong-field
surface. At $i=20^\circ$, the inner-born intensity fractions rise to $10\%$
and $6.4\%$, while final polarizing surfaces inside $20r_s$ contribute $20\%$
and $16\%$ of the absolute $Q$ terms. Azimuthal cancellation nevertheless
keeps the net polarization below $0.03\%$ at both wavelengths. Relativistic
corrections may weakly affect this small low-inclination residual, but are
negligible at the relevant inclinations: all observed polarization arises at
weak-field surfaces and inner-born photons contribute below $1.1\%$ of the
intensity.

\begin{figure}[!tbp]
\centering
\setcounter{section}{4}
\setcounter{figure}{0}
\includegraphics[width=\columnwidth]{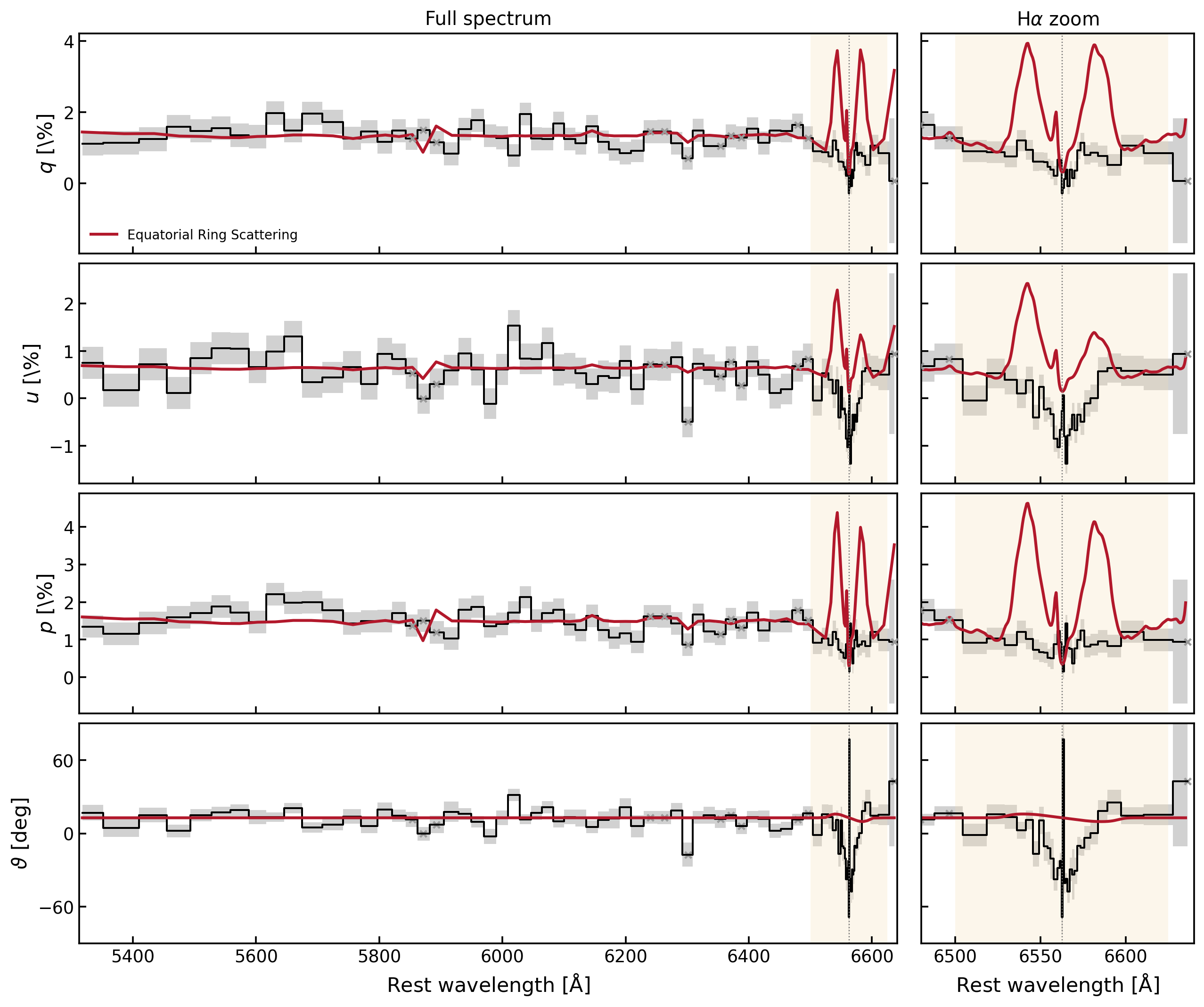}
\caption{Equatorial ring-scattering model for J1025+1402. Black
histograms and gray bands show the data and $1\sigma$ uncertainties; red
curves show the best-fitting model. The continuum sets the fitted amplitude
and $\psi_{\rm axis}$, leaving the H$\alpha$ Stokes profiles as predictions.
The displayed profiles are $q=Q/I_{\rm obs}$ and $u=U/I_{\rm obs}$, whereas
the likelihood is evaluated in $Q$ and $U$ flux.
The model reproduces the continuum but predicts excessive broad-line
structure. The fit has nominal
$\chi^2/{\rm dof}=1056/133=7.94$.}
\label{fig:equatorial_best}
\end{figure}

\begin{figure}[!tbp]
\centering
\setcounter{section}{5}
\setcounter{figure}{0}
\includegraphics[width=\columnwidth]{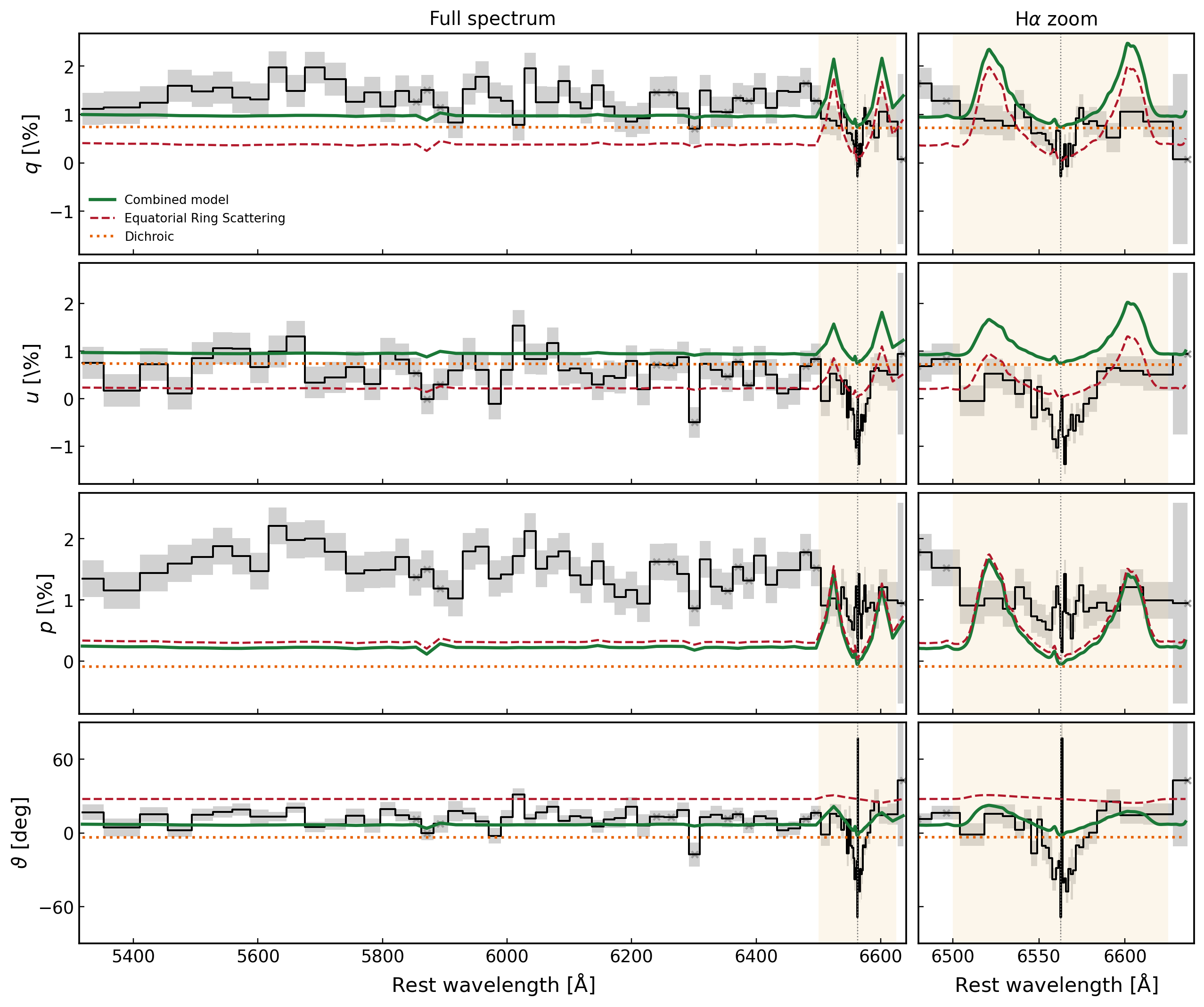}
\caption{Equatorial ring scattering plus dichroic transmission for
J1025+1402. Black histograms and gray bands show the data and $1\sigma$
uncertainties. Green curves show the combined model, while red dashed and
orange dotted curves show the equatorial-scattering and dichroic components,
respectively. The displayed profiles are normalized $q$ and $u$; the
likelihood uses $Q$ and $U$ flux. Partial cancellation reduces the
scattered-line signal, but the
model still underpredicts the continuum polarization and overpredicts the
broad-line Stokes structure. The fit has nominal
$\chi^2/{\rm dof}=450/131=3.43$.}
\label{fig:equatorial_dichroic}
\end{figure}

\begin{figure}[!tbp]
\centering
\setcounter{section}{6}
\setcounter{figure}{0}
\includegraphics[width=\columnwidth]{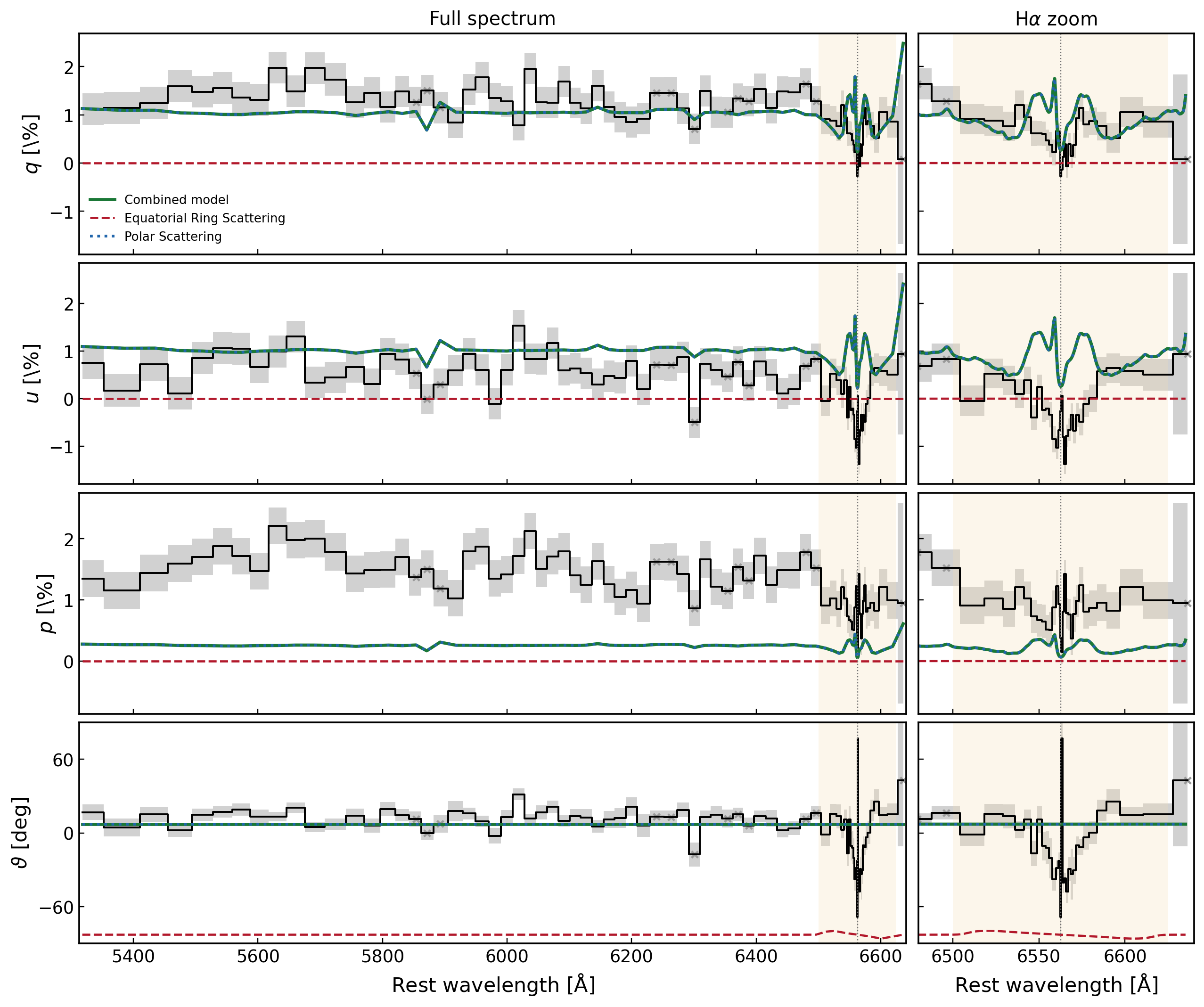}
\caption{Equatorial ring plus polar-cone scattering for
J1025+1402. Black histograms and gray bands show the data and $1\sigma$
uncertainties. Green curves show the combined model, while blue dotted curves
show the polar component. The best fit drives the equatorial contribution to
zero, so that component is not plotted; the remaining polar model
underpredicts the continuum polarization while producing excessive broad-line
Stokes structure. The displayed profiles are normalized $q$ and $u$; the
likelihood uses $Q$ and $U$ flux. The fit has nominal
$\chi^2/{\rm dof}=461/130=3.54$.}
\label{fig:equatorial_polar}
\end{figure}

\begin{figure}[!tbp]
\centering
\setcounter{section}{7}
\setcounter{figure}{0}
\includegraphics[width=\columnwidth]{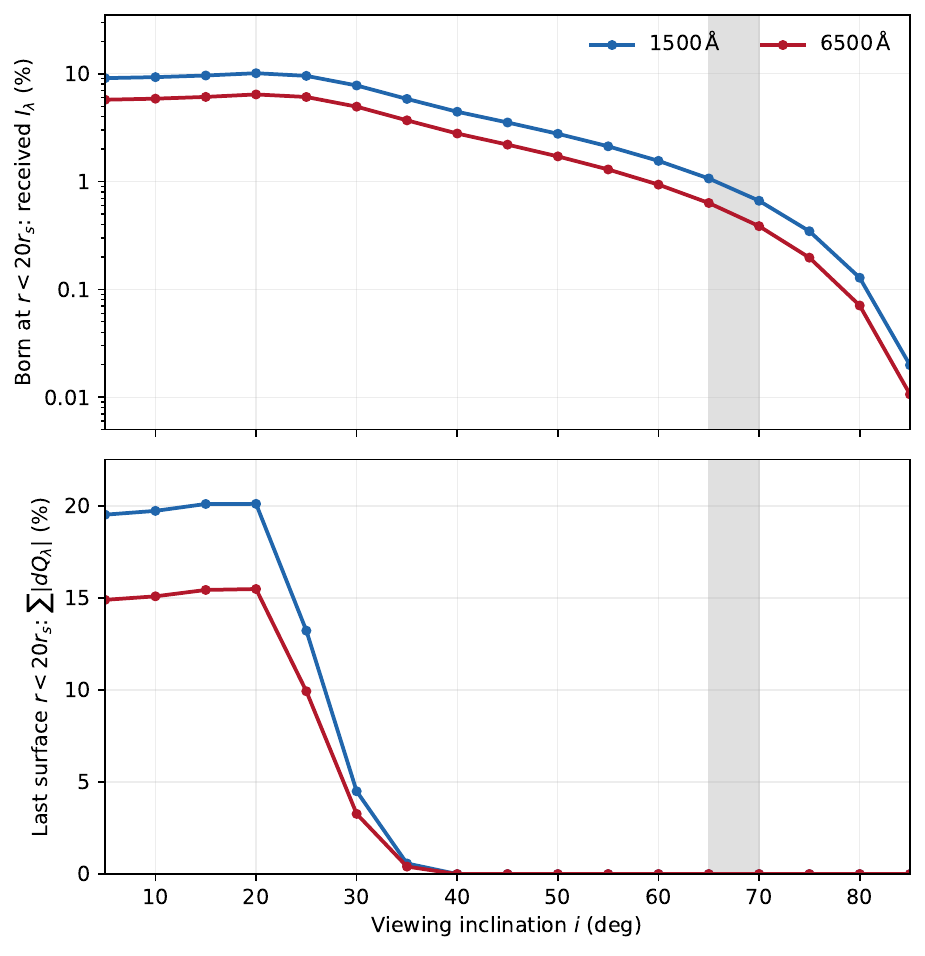}
\caption{Strong-field diagnostics for the
$M_{\rm BH}=10^{7.5}\,M_\odot$, $\dot m=31.7$ model. The upper panel is the
fraction of received intensity born at $r<20r_s$, including all subsequent
scalar wall reflections. The lower panel is the fraction of
$\sum|\mathrm dQ|$ whose final polarizing surface lies at $r<20r_s$. Blue and
red denote rest-frame $1500\,$\AA\ and $6500\,$\AA; the shaded band marks
$i=65^\circ$--$70^\circ$. Both panels use decimal percentage notation; the
upper axis is logarithmic and the lower axis is linear.}
\label{fig:gr_origin_fraction}
\end{figure}

\end{appendix}

\end{document}